\documentclass[aps,floats]{revtex4}
\usepackage{amsmath,amssymb}
\usepackage{graphicx,epsfig}

\begin{document}
\bibliographystyle {plain}

\def\oppropto{\mathop{\propto}} 
\def\opsimeq{\mathop{\simeq}}
\def\opoverderline{\mathop{\overline}}
\def\operarrow{\mathop{\longrightarrow}}
\def\opsim{\mathop{\sim}}

\def\fig#1#2{\includegraphics[height=#1]{#2}}
\def\figx#1#2{\includegraphics[width=#1]{#2}}


\title{ Non-equilibrium dynamics of finite-dimensional disordered systems :
 \\ RG flow towards an ``infinite disorder'' fixed point at large times.
 } 


 \author{ C\'ecile Monthus and Thomas Garel }
  \affiliation{Institut de Physique Th\'{e}orique, CNRS and 
CEA Saclay, 91191 Gif-sur-Yvette cedex, France}

\begin{abstract}

To describe the non-equilibrium dynamics of random systems, we have recently introduced (C. Monthus and T. Garel, arxiv:0802.2502) a 'strong disorder renormalization' (RG) procedure in configuration space that can be defined for any master equation. In the present paper, we analyze in details the properties of the large time dynamics whenever the RG flow is towards some ``infinite disorder'' fixed point, where the width of the renormalized barriers distribution grows indefinitely upon iteration. In particular, we show how the strong disorder RG rules can be then simplified while keeping their asymptotic exactness, because the preferred exit channel out of a given renormalized valley typically dominates asymptotically over the other exit channels. We explain why the present approach is an explicit construction in favor of the droplet scaling picture where the dynamics is governed by the logarithmic growth of the coherence length $l(t) \sim \left( \ln t \right)^{1/\psi}$, and where the statistics of barriers corresponds to a very strong hierarchy of valleys within valleys. As an example of application, we have followed numerically the RG flow for the case of a directed polymer in a two-dimensional random medium. The full RG rules are used to check that the RG flow is towards some infinite disorder fixed point, whereas the simplified RG rules allow to study bigger sizes and to estimate the barrier exponent $\psi$ of the fixed point.

\end{abstract}

\maketitle

\section{ Introduction   }

The non-equilibrium dynamics of disordered systems gives rise to
a lot of striking properties that have been much studied both theoretically
and experimentally (see \cite{bouchaud,bouchaudetal,berthierhouches}
 and references therein).
The first property that has attracted a lot of attention is 'aging' 
which can be seen as some 'criticality in the time direction', 
in the following sense : if the dynamics has taken place during
the time interval $[0,t_w]$, the only relevant time scale for the dynamics
at larger times $t>t_w$ is the time $t_w$ itself.
A well known example is the case of phase ordering of finite-dimensional
systems when the dynamics tends towards equilibrium
presenting some long-ranged order. It is then useful to introduce
some coherence length $l(t)$
that separates the smaller lengths $l < l(t)$ which are quasi-equilibrated
from the bigger lengths $l > l(t)$ which are completely
out of equilibrium. Then equilibrium is reached only when
the coherence length reaches the macroscopic linear
size $l(t_{eq}) = L$ of the system. 
In pure systems, these phenomena of phase ordering are well understood
 \cite{brayreview} and the coherence length grows algebraically
\begin{eqnarray}
 l_{pure}(t) \sim t^{1/z}
\label{lpure}
\end{eqnarray}
with some dynamical exponent $z$ \cite{brayreview}.
It is important to note that for pure systems, domain growth
is possible even at zero-temperature because 
domain walls can still move and annihilate.
 In the presence of quenched disorder however, 
the dynamics requires thermal activation (in particular,
at zero temperature, the dynamics 
stops on the first encountered local minimum).
Within the droplet scaling theory
proposed both for spin-glasses \cite{heidelberg,Fis_Hus} 
and for directed polymers in 
random media \cite{Fis_Hus_DP}, the barriers grow 
as a power law of the length $l$
\begin{eqnarray}
B(l) \sim l^{\psi}
\label{defpsi}
\end{eqnarray}
with some barrier exponent $\psi>0$.
The typical time $t_{typ}(l)$ associated to scale $l$ grows as an exponential
$\ln t_{typ}(l) \sim B(l) \sim l^{\psi}$. 
Equivalently, the characteristic
length-scale $l(t)$ associated to time $t$ grows only logarithmically in time
\begin{eqnarray}
l(t) \sim \left( \ln t \right)^{\frac{1}{\psi}}
\label{typltime}
\end{eqnarray}
In numerical studies, this logarithmic behavior has
 remained controversial, because the dynamics is very
slow not only in real life but also in Monte-Carlo simulations!
As a consequence, the maximal equilibrated length $l_{max} $ measured
 at the end of the simulations is usually rather small,
so that various fits of the data are possible.
Some authors use the algebraic fit of Eq. \ref{lpure}
with a temperature and disorder dependent exponent $z(T,\epsilon)$
either for disordered ferromagnets \cite{rieger_coarsening}
or for spin-glasses \cite{kis,kat_cam},
whereas logarithmic fits corresponding to Eq. \ref{typltime}
can be found in \cite{Hus_Hen,puri,bray_hum}
for disordered ferromagnets and in \cite{huseSG,berthier} for spin-glasses.
For the case of an elastic line in a random medium,
various authors have also used algebraic time scalings
to fit aging data \cite{DP_alge}, but more recently
Kolton, Rosso and Giamarchi \cite{rosso}
have been able to exclude the
power-law $l(t) \sim t^{1/z}$ at large scales and to measure
a barrier exponent $\psi>0$ in Eq. \ref{typltime} which is asymptotically 
size and time independent as it should. 
However, since fits of numerical data in most disordered systems
will probably remain controversial for a long time,
we feel that more detailed theoretical arguments should be
provided in favour of either algebraic or logarithmic behavior.
In this paper, we explain why the strong disorder renormalization (RG)
approach in configuration space introduced recently 
\cite{letter} is an explicit construction in favor of the 
droplet logarithmic scaling of Eq. \ref{typltime}.  

Besides aging properties of disordered systems at a given temperature, 
physicists have been also interested into
more complicated temperature cycling experiments 
that display rejuvenation and memory (see
 \cite{bouchaud,bouchaudetal,berthierhouches} and references therein
for more details). The important point for the present discussion 
is that these phenomena require some hierarchical organization 
of valleys within valleys, where the rejuvenation due to short length scales
does not destroy the memory of large length scales
 which are effectively frozen. Since this hierarchy is sometimes believed
to be present only in mean-field models, we would like to stress here
that the droplet logarithmic scaling of Eq. \ref{typltime}
effectively leads to a clear separation of time scales
and to an effective hierarchy of valleys at large scales,
as already argued in \cite{bouchaud,bouchaudetal}. 
The strong disorder RG procedure 
that we discuss in the present paper is in full agreement with these ideas,
since we expect that for a very broad class of disordered systems in their
glassy phase, the RG procedure 
flows towards some ``infinite disorder fixed point''
that precisely describes a strong hierarchy of valleys within valleys.

The paper is organized as follows.
In Section \ref{rgconfig}, we recall the principles of 
strong disorder renormalization in configuration space
 introduced in \cite{letter} and discuss its properties.
In Section \ref{numefull}, we follow numerically the RG flow
corresponding to the directed polymer in a two dimensional
random medium
and find evidence of convergence towards an ``infinite disorder'' fixed point.
In Section \ref{simpliRG}, we introduce
 simplified RG rules  
near ``infinite disorder'' fixed point.
In Section \ref{numesimpli},
 we present the numerical results based on simplified RG rules
that allow to study bigger sizes and to estimate the barrier exponent 
$\psi$ of the fixed point.
In section \ref{secpsi}, we discuss the physical meaning
of the barrier exponent $\psi$ for the structure of 
renormalized valleys in the configuration space.
Our conclusions are summarized in Section \ref{conclusion}.

\section{ Strong disorder RG rules for 
random master equations } 

\label{rgconfig}


Strong disorder renormalization 
(see \cite{review} for a review) is a very specific type of RG
that has been first developed in the field of quantum spins :
the RG rules of Ma and Dasgupta \cite{madasgupta} 
have been put on a firm ground by D.S. Fisher 
who introduced the crucial idea of ``infinite disorder'' fixed point
where the method becomes asymptotically exact,
and who computed explicitly exact 
critical exponents and scaling functions 
for one-dimensional disordered quantum spin chains \cite{dsf}.
This method has thus generated a lot of activity for various
disordered quantum models \cite{review}, and has been then
successfully applied to
various classical disordered dynamical models,
such as random walks in random media \cite{sinairg,sinaibiasdirectedtraprg},
reaction-diffusion in a random medium \cite{readiffrg}, 
coarsening dynamics of classical spin chains \cite{rfimrg}, 
trap models \cite{traprg}, random vibrational networks \cite{vibrational},
absorbing state phase transitions \cite{contactrg},
zero range processes \cite{zerorangerg} and 
exclusion processes  \cite{exclusionrg}.
In all these cases, the strong disorder RG rules 
have been formulated {\it in real space},
with specific rules depending on the problem.
For more complex systems where
 the formulation of strong disorder RG rules
has not been possible in real space, 
we have recently proposed in \cite{letter} a strong disorder 
RG procedure { \it in configuration space} that can be
 defined for any master equation. In the remaining of this section,
we describe this procedure and discuss its properties in more details.

\subsection{ Master equation defining the dynamics} 

In statistical physics, it is convenient to define the dynamics via a
 Master Equation describing the evolution of the
probability $P_t ({\cal C} ) $ to be in a configuration ${\cal C}$ at time t
\begin{eqnarray}
\frac{ dP_t \left({\cal C} \right) }{dt}
= \sum_{\cal C '} P_t \left({\cal C}' \right)
 W \left({\cal C}' \to  {\cal C}  \right) 
 -  P_t \left({\cal C} \right) W_{out} \left( {\cal C} \right)
\label{master}
\end{eqnarray}
The notation  
$ W \left({\cal C}' \to  {\cal C}  \right) $ 
represents the transition rate per unit time from configuration 
${\cal C}'$ to ${\cal C}$, and the notation
\begin{eqnarray}
W_{out} \left( {\cal C} \right)  \equiv
 \sum_{ {\cal C} '} W \left({\cal C} \to  {\cal C}' \right) 
\label{wcout}
\end{eqnarray}
represents the total exit rate out of configuration ${\cal C}$.
The two important properties of this master equation are the following :

(i)  the exit time $\tau$ from configuration  ${\cal C}$
is a random variable distributed with the law
\begin{eqnarray}
P^{exit}_{\cal C} (\tau) = W_{out} \left( {\cal C} \right)
 e^{ -  \tau W_{out} \left( {\cal C} \right)} 
\label{pexit}
\end{eqnarray}
with the normalization $\int_0^{+\infty} d\tau P^{exit}_{\cal C} (\tau)=1$.

(ii) the new configuration ${\cal C}'$
where the system jumps at time $\tau$
when it leaves the configuration ${\cal C}$
is chosen with the probability
\begin{eqnarray}
\pi_{\cal C} \left({\cal C}' \right) = 
\frac{W \left({\cal C}  \to {\cal C}' \right)}{W_{out} \left( {\cal C} \right)}
\label{pconfigjump}
\end{eqnarray}
normalized to $\sum_{\cal C '}\pi_{\cal C} \left({\cal C}' \right)=1$.

These two properties are the basis of faster-than-the-clock algorithms, called 
 'Bortz-Kalos-Lebowitz algorithm' \cite{algoBKL} in physics (and
 'Gillespie algorithm' \cite{gillespie} in chemistry),
where each iteration leads to a movement. 
However, even if these algorithms avoid trapping in a given microscopic
configuration, they do not avoid trapping in a valley of configurations.
As a consequence, these algorithms which are usually very
powerful for pure systems at low temperature
become inefficient in the presence of frozen disorder because they face the 'futility' problem \cite{werner} : the number of distinct configurations visited during the simulation remains very small 
with respect to the accepted moves. The reason is that the system
 visits over and over again the same configurations
 within a given valley before it is able to
escape towards another valley. 
This is why we propose in the following some renormalization procedure that allows to work directly with the 'valleys' of configurations
on larger and larger time scales.

\subsection{ Statement of the strong disorder renormalization rules }

For dynamical models, the aim of any renormalization procedure
is to integrate over 'fast' processes to obtain effective properties 
of 'slow' processes.
 The general idea of 'strong renormalization' for dynamical models
consists in eliminating iteratively the 'fastest' process.
The RG procedure introduced
in \cite{letter} can be summarized as follows :

(1) find the configuration ${\cal C}^*$ with the biggest exit rate $W^*_{out}$
(i.e. the smallest exit time, see Eq. \ref{pexit}) 
\begin{eqnarray}
W^*_{out} = W_{out} \left( {\cal C}^* \right)
 \equiv  {\rm max}_{{\cal C}} \left[  W_{out} \left( {\cal C} \right) \right]
\label{defwmax}
\end{eqnarray}

(2) find the neighbors $({\cal C}_1,{\cal C}_2,...,{\cal C}_n)$
 of configuration ${\cal C}^*$, 
i.e. the configurations that were related 
via positive rates $W({\cal C}^* \to {\cal C}_i )>0$ and
 $W({\cal C}_i\to {\cal C}^*)>0$
to the decimated configuration ${\cal C}^*$
(here we will assume for the simplicity of the discussion,
 and because it is usually the case in
statistical physics models, that if a transition has 
a strictly positive rate, the reverse transition has also
a strictly positive rate; but of course
 the renormalization rules can be simply extended to other cases).
For each neighbor configuration ${\cal C}_i$ with
$i \in (1,..,n)$, update the transition rate to go to
the configuration ${\cal C}_j$ with $j \in (1,..,n)$ and $j \neq i$
 according to
\begin{eqnarray}
W^{new}({\cal C}_i \to {\cal C}_j )=W({\cal C}_i \to {\cal C}_j )
+ W({\cal C}_i \to {\cal C}^* ) \times  \pi_{{\cal C}^*} \left({\cal C}_j \right) 
\label{wijnew}
\end{eqnarray}
where the first term represents the 'old' transition rate (possibly zero),
and the second term represents the transition 
via the decimated configuration ${\cal C}^*$ :
the factor $W({\cal C}_i \to {\cal C}^* ) $ takes into account 
the transition rate to ${\cal C}^*$ and the term
\begin{eqnarray}
\pi_{{\cal C}^*} \left({\cal C}_j \right) = 
\frac{W \left({\cal C}^*  \to {\cal C}_j \right)}{W_{out} \left( {\cal C}^*\right)}
\label{picstar}
\end{eqnarray}
represents the probability
 to make a transition towards ${\cal C}_j$
when in ${\cal C}^*$ (see Eq. \ref{pconfigjump}).
The $2 n$ rates $W({\cal C}^* \to {\cal C}_i )$
 and $W({\cal C}_i \to {\cal C}^*)$ then
 disappear with the decimated configuration ${\cal C}^*$.
Note that the rule of Eq. \ref{wijnew} 
has been recently proposed in \cite{vulpiani}
to eliminate 'fast states'  from various dynamical problems 
with two very separated time scales.
The physical interpretation of this rule is as follows :
the time spent in the decimated configuration ${\cal C}^*$ is neglected
with respects to the other time scales remaining in the system. 
The validity of this approximation within the present renormalization procedure
will be discussed in detail below. 

(3) update the exit rates out of the neighbors ${\cal C}_i$, with $i=1,..,n$
either with
the definition
\begin{eqnarray}
W^{new}_{out}({\cal C}_i)   = 
\sum_{\cal C} W^{new}({\cal C}_i \to {\cal C} )
\label{wioutnewactualisation}
\end{eqnarray}
or with the rule that can be deduced from Eq. \ref{wijnew}
 \begin{eqnarray}
W^{new}_{out}({\cal C}_i) = W_{out}({\cal C}_i) 
 - W({\cal C}_i \to {\cal C}^* ) 
\frac{ W({\cal C}^* \to {\cal C}_i )}
{W^{*}_{out}}
\label{wioutnew}
\end{eqnarray}
(since this rule contains a subtraction, it can be used numerically
only with great care !).
The physical meaning of this rule is the following.
The exit rate out of the configuration ${\cal C}_i$ decays because 
the previous transition towards ${\cal C}^*$ can lead to an immediate return
towards ${\cal C}_i$ with probability 
$\pi_{{\cal C}^*} \left({\cal C}_i\right) =
\frac{ W({\cal C}^* \to {\cal C}_i )}
{W^{*}_{out}} $. After the decimation of the configuration ${\cal C}^*$,
this process is not considered as an 'exit' process anymore, but as a
residence process in the configuration ${\cal C}_i$.
This point is very important to understand the
 meaning of the renormalization procedure :
the remaining configurations at a given renormalization scale are 
'formally' microscopic configurations 
of the initial master equation (Eq. \ref{master}),
but each of these remaining microscopic configuration
 actually represents some 'valley' in configuration space
that takes into account all the previously decimated configurations.

(4) return to point (1).

Note that in practice, the renormalized rates $W({\cal C} \to {\cal C}' )$
can rapidly  become very small as a consequence of the multiplicative structure
of the renormalization rule of Eq \ref{wijnew}. This means that the appropriate
 variables are the logarithms of the transition rates, that we will call 'barriers' in the remaining
of this paper. The barrier $B ({\cal C} \to {\cal C}' )$ from ${\cal C}$  to ${\cal C}' $ is defined by
\begin{eqnarray}
B ({\cal C} \to {\cal C}' )\equiv - \ln W({\cal C} \to {\cal C}' )
\label{defb}
\end{eqnarray}
and similarly the exit barrier out of configuration ${\cal C}$ is defined by
\begin{eqnarray}
B_{out} ({\cal C} )\equiv - \ln W_{out}({\cal C} )
\label{defbout}
\end{eqnarray}
Note that a very important advantage of this formulation in terms
of the renormalized transition rates of the master equation is that 
the renormalized barriers take into account the true 'barriers'
of the dynamics, whatever their origin which can be
 either energetic or entropic.

\subsection{ Notion of 'infinite disorder fixed point' 
and asymptotic exactness of the RG rules  }

\label{strongdisorderfixedpoint}

As mentioned above, the approximation made
 in the renormalization rule of Eq. \ref{wijnew}
consists in neglecting the time spent 
in the decimated configuration ${\cal C}^*$ 
with respect to the other time scales remaining in the system. 
In the present framework, this means that the maximal exit rate chosen
 in Eq \ref{defwmax}
should be well separated from the exit rates of the neighboring
 configurations ${\cal C}_i$.
The crucial idea of 'infinite disorder fixed point' \cite{dsf,review}
is that even if this approximation is not perfect during the
 first steps of the renormalization,
this approximation will become better and better 
 at large time scale if the probability distribution
 of the remaining exit rates
becomes broader and broader upon iteration. 
More precisely,  if the renormalization scale $\Gamma$ 
is defined as the exit barrier of the
 last eliminated configuration ${\cal C^*}$
\begin{eqnarray}
\Gamma= B_{out} ({\cal C^*} )\equiv - \ln W_{out}^*
\label{defgamma}
\end{eqnarray}
one expects that the probability distribution
 of the remaining exit barrier $B_{out} \geq \Gamma$ will 
converge towards some scaling form
\begin{eqnarray}
P_{\Gamma} ( B_{out} -\Gamma ) \opsimeq_{ \Gamma \to \infty} 
  \frac{1}{\sigma(\Gamma) } {\hat P} 
\left( \frac{B_{out} - \Gamma}{\sigma(\Gamma) } \right)
\label{pgammabout}
\end{eqnarray}
where ${\hat P} $ is the fixed point probability distribution, 
and where $\sigma(\Gamma)$ is the appropriate scaling factor
for the width.
The notion of 'infinite disorder fixed point' 
means that the width $\sigma(\Gamma)$ grows indefinitely with the
renormalization scale $\Gamma$
\begin{eqnarray}
\sigma(\Gamma) \opsimeq_{ \Gamma \to \infty} +\infty
\label{infinite}
\end{eqnarray}
Whenever this 'infinite disorder fixed point' condition 
 is satisfied, the strong disorder renormalization
procedure becomes asymptotically exact at large scales.
In previously known cases of infinite disorder fixed
 points where calculations can be done explicitly
\cite{review},  the scale $\sigma(\Gamma)$ has been 
found to grow linearly 
\begin{eqnarray}
\sigma(\Gamma) \opsimeq_{ \Gamma \to \infty} \Gamma
\label{sigmalinear}
\end{eqnarray}
This behavior means that the cut-off $\Gamma$ is the
 only characteristic scale and thus describes some critical 
point \cite{review}.
For the present procedure concerning the dynamics in disordered models,
this property means some 'criticality in the time direction',
i.e. the absence of any characteristic time scale between the microscopic scale
and the macroscopic equilibrium time of the full disordered sample.
As explained at the beginning of the introduction,
this 'criticality in the time direction' will naturally leads to 
aging behaviors for two-time properties. An example where 
asymptotically exact two-time aging properties
 have been explicitly computed via strong disorder RG
is the Sinai model \cite{sinairg}.   

When the width $\sigma(\Gamma)$ instead converges towards
 a finite value $\sigma(\infty)<+\infty$,  
one speaks of a `finite-disorder fixed point'.
However, if this constant $\sigma(\infty)$ is large,
 one speaks of a 'strong disorder fixed point', and the validity
of the RG approach is of order $1/\sigma(\infty)$ : we refer to
\cite{sinaibiasdirectedtraprg} where systematic expansions
 in $1/\sigma(\infty)$ with respect to the leading strong disorder
RG have been explicitly computed.
This notion of 'strong disorder fixed point'
is very useful to study the vicinity of 
'infinite disorder fixed point' in the space of parameters \cite{review}.
For instance in the Sinai model, the 
'infinite disorder fixed point' is realized in the absence of drift
where the diffusion is logarithmically slow, 
whereas the 'strong disorder fixed point' 
corresponds to the presence of a small drift where 
the diffusion is algebraic but with an anomalous exponent 
\cite{sinairg,sinaibiasdirectedtraprg}.

For the present strong disorder renormalization of a master equation,
the convergence towards an 'infinite disorder fixed point'  
will depend on the initial condition of the transition rates,
i.e. on the model (and on the temperature if there are phase transitions).
However, the form of the RG rules of Eq \ref{wijnew}
 is sufficiently similar to the usual Ma-Dasgupta rules
 \cite{review} to think that 
 the convergence towards some infinite disorder fixed point
 should be realized in
 a very broad class of disordered systems in their glassy phase.
In practice, it should be checked numerically for each model of interest.

\section{ Numerical studies of the full RG procedure }

\label{numefull}

\subsection{ Main numerical limitation : proliferation of neighbors }

\label{proliferation}

In dimension $d=1$, strong disorder RG rules maintain
the one-dimensional structure where each site has two neighbors,
one on the left and one on the right, and this is why one can obtain
explicit exact solutions \cite{review}. 
In dimension $d>1$, strong disorder RG rules
change the local coordination numbers and usually lead to a 
 proliferation of neighbors
as already found in real-space strong disorder RG studies of quantum models 
 \cite{motrunich,lin}.
With the present notations, the reason is clear 
from the RG rule of Eq. \ref{wijnew} : if the decimated configuration
${\cal C}^*$ has $n$ neighbors $({\cal C}_1,{\cal C}_2,...,{\cal C}_n)$, 
one eliminates $(2n)$ rates 
(the rates $W({\cal C}^* \to {\cal C}_i )$ and
 $W({\cal C}_i\to {\cal C}^*)$ for $i=1,2,..n$) but one can create up to
$n (n-1)$ transition rates (the rates $W({\cal C}_i \to {\cal C}_j )$
with  $i=1,2,..n$ and $j \ne i$).
The increase in the total number $N^{rates}$ of transitions rates
when one decimates a configuration ${\cal C}^*$ with $n$ neighbors
is thus only bounded by
\begin{eqnarray}
\Delta N^{rates} \leq n(n-3)
\label{deltanrates}
\end{eqnarray}
In particular, each of the $n$ neighbors ${\cal C}_i$
looses one neighbor (${\cal C}^* $), but can gain up to
$(n-1)$ new neighbors, so that the increase of its
coordination number $z$ is only bounded by
\begin{eqnarray}
\Delta z \leq n-2
\label{deltaz}
\end{eqnarray}
For an initial  master equation describing local single moves, 
the first applications of the strong disorder RG procedure 
will establish new links between configurations that were
not initially related via single moves.
As a consequence, the number $N^{rates}$ of rates
and the coordination $z$ of the surviving configurations
will increase during the first stages of the renormalization
to describe moves made of two,three,.. elementary moves.

In real-space strong disorder RG studies of quantum models with couplings
$J_{ij}$, a numerical cut-off $J_{min}$ is usually introduced to keep the new
interactions only if they are above the cut-off  $J_{ij}>J_{min}$,
whereas weaker bonds $J_{ij}<J_{min}$ are disregarded \cite{motrunich}.
Within the present framework where transition rates are not symmetric
($W({\cal C}_i \to {\cal C}_j ) \ne W({\cal C}_j \to {\cal C}_i )$)
and where the renormalization concerns the exit rates out of surviving
configurations, the problem of simplifying the RG rules numerically
is different. In the next section \ref{simpliRG}, we will
propose simplified RG rules that are valid
 near ``infinite disorder'' fixed points.
But before studying these simplified RG flows,
it is important to check that the full RG flow
starting from an initial condition describing
 the dynamical models of interest
indeed flows towards some ``infinite disorder'' fixed point.
In the remaining of this section, 
we thus study numerically the full RG flow
for the special case of a directed polymer in a two dimensional random medium.

\subsection{ Example : directed polymer in a two dimensional random medium }

\label{DPmodel}

We consider a directed polymer of length $L$ with a fixed origin :
the ${\cal N}_0 = 2^L$ possible configurations are 
given by the sequence of heights $(h_1,h_2,...,h_L)$ 
that satisfy the chain constraints 
\begin{eqnarray}
h_x- h_{x-1} = \pm 1
\label{chain}
\end{eqnarray}
for $x=1,2,..,L$  with the boundary condition $h_0=0$.
The energies of these configurations are given by
\begin{eqnarray}
E \left( {\cal C} = (h_1,h_2,...,h_L) \right) = \sum_{x=1}^L \epsilon (x,h_x)
\label{defEc}
\end{eqnarray}
where the site random energies 
$\epsilon (i,h)$ are frozen variables that represent the random medium.
We consider the case where these energies are independent and
drawn from the Gaussian distribution 
\begin{eqnarray}
\rho( \epsilon) = \frac{1}{\sqrt{2 \pi}} e^{-\epsilon^2/2}
\label{epsgauss}
\end{eqnarray}

For the directed polymer model, we are interested into 
the local Metropolis dynamics
defined by the transition rates 
\begin{eqnarray}
W \left( {\cal C} \to {\cal C}'  \right)
= \delta_{<{\cal C}, {\cal C}' >} 
\  {\rm min} \left(1, e^{-  (E({\cal C}' )-E({\cal C} ))/T } \right)
\label{metropolis}
\end{eqnarray}
The first factor $\delta_{<{\cal C}, {\cal C}' >}$
 means that the two configurations
are related by a single move, and
 the last factor ensures the convergence towards thermal equilibrium
at temperature $T$ via the detailed balance property
\begin{eqnarray}
 e^{- E({\cal C}) /T} W \left( {\cal C} \to {\cal C}'  \right)
= e^{- E({\cal C}') /T} W \left( {\cal C}' \to {\cal C}  \right)
\label{detailedbalance}
\end{eqnarray}
In contrast with spin models, where a configurations of $N$ spins
is related to exactly $N$ other configurations by single flips,
a configuration of $L$ monomers of the directed polymer is
usually not related to $L$ other configurations as a consequence
of the chain constraints of Eq. \ref{chain}.
More precisely, if we call configuration ${\cal C}^{[x]\pm}$ the configuration
obtained from  ${\cal C}$ by the elementary move $h_x \to h_x \pm 2$,
we note that, as a consequence of the chain constraint, 
this elementary move is possible only if 
the two neighbors are in the favorable positions $h_{x \pm 1} = h_x \pm 1$.
The energy change associated to the elementary move $h_x \to h_x \pm 2$
reads in terms of the random energies introduced in Eq. \ref{defEc} 
\begin{eqnarray}
E({\cal C} ^{[x]\pm} ) -E({\cal C} )
 \equiv \epsilon (x,h_x \pm 2) - \epsilon (x,h_x)
\label{deltaE}
\end{eqnarray}

With the full RG rules, where the problem of proliferation of neighbors
discussed in section \ref{proliferation} is memory and time consuming, 
the linear sizes we have been able to study
are rather small $ L \leq 11$ (the number 
 of configurations grows exponentially $ 2^L \leq 2048$).
The corresponding numbers $n_s(L)$ 
of disordered samples of length $L$ read 
\begin{eqnarray}
 L && =  5, 6, 7, 8, 9, 10, 11  \nonumber \\
n_s(L) && = 7.10^6, 10^6, 10^5, 14.10^5, 15.10^4, 14.10^3, 500
\label{sizesfullRG}
\end{eqnarray}

\subsection{ Analysis of the numerical data : two useful ensembles  }

To analyze the numerical results concerning the application of
the strong disorder renormalization to many disordered 
samples, it is interesting to consider two different 'ensembles'
corresponding to two types of averages as we now explain.

\subsubsection{ Averaging over disordered 
samples at fixed RG scale $\Gamma$  }

\label{fixedgamma}

The first ensemble consists in collecting data at fixed RG scale $\Gamma$,
where $\Gamma$ is the last decimated
renormalized exit barrier remaining in the system.
The advantage is that the comparison with theoretical statements
is more straightforward, since many of the statements concern
a fixed RG scale, in particular the probability distribution of Eq.
\ref{pgammabout} that defines the scaling properties of the barriers.
However, since the RG scale $\Gamma$ is a continuous variable,
one needs then to introduce 
some appropriate discretization numerically.
For instance in the results presented below, we have chosen
a window of width $\Delta \Gamma = 0.1$.
In conclusion,  ``data at fixed RG scale $\Gamma$'' 
correspond to an average over the disordered samples where 
the last decimated renormalized exit barrier remaining in the system is
within a window of width $\Delta \Gamma = 0.1$ around $\Gamma$.

\subsubsection{ Averaging over disordered 
samples at fixed number ${ \cal N}$ of surviving configurations
(i.e. at fixed coherence length) }

\label{fixednconfig}

However another way of analyzing numerical data, used for instance
in strong disorder RG study of quantum models in dimension $d>1$
\cite{motrunich} consists in collecting data corresponding
to a fixed number of RG steps, or equivalently to a fixed number
${ \cal N}$ of surviving configurations. 
This type of averaging can be considered as a fixed-length ensemble
as we now explain.
In strong disorder RG study of quantum models in dimension $d>1$
\cite{motrunich}, the number $N$ of surviving spins 
 when starting from $N_0=L^d$ initial spins
can be used to define a length scale
$l$ via $N = \left( L/l \right)^d$.
For the present renormalization where 
the number of initial configurations is ${ \cal N}_0=2^{L}$,
we may similarly define a length $l$
via
\begin{eqnarray}
 { \cal N} \equiv  2^{\frac{L}{l}}
\label{ngamma}
\end{eqnarray}
This length $l$ represents some 
some growing correlation length in the following sense :
each segment of length $l$ of the polymer 
corresponds to one renormalized degree of freedom. 
Initially this length is one and corresponds to a single monomer,
whereas at the end of the renormalization process where equilibrium
is reached, the
number of configurations reaches $ { \cal N}_{eq}=1$ and
the coherence length reaches the total length $l_{eq}=L$
of the polymer.
So besides the studies at fixed RG scale $\Gamma$ described above, 
it is also interesting to consider ``data at fixed ${ \cal N}$''
as in \cite{motrunich}
where data are averaged over disordered samples
having the same number 
${ \cal N}$ of surviving renormalized configurations.

\subsection{Probability distribution of the renormalized exit barriers }

\begin{figure}[htbp]
\includegraphics[height=6cm]{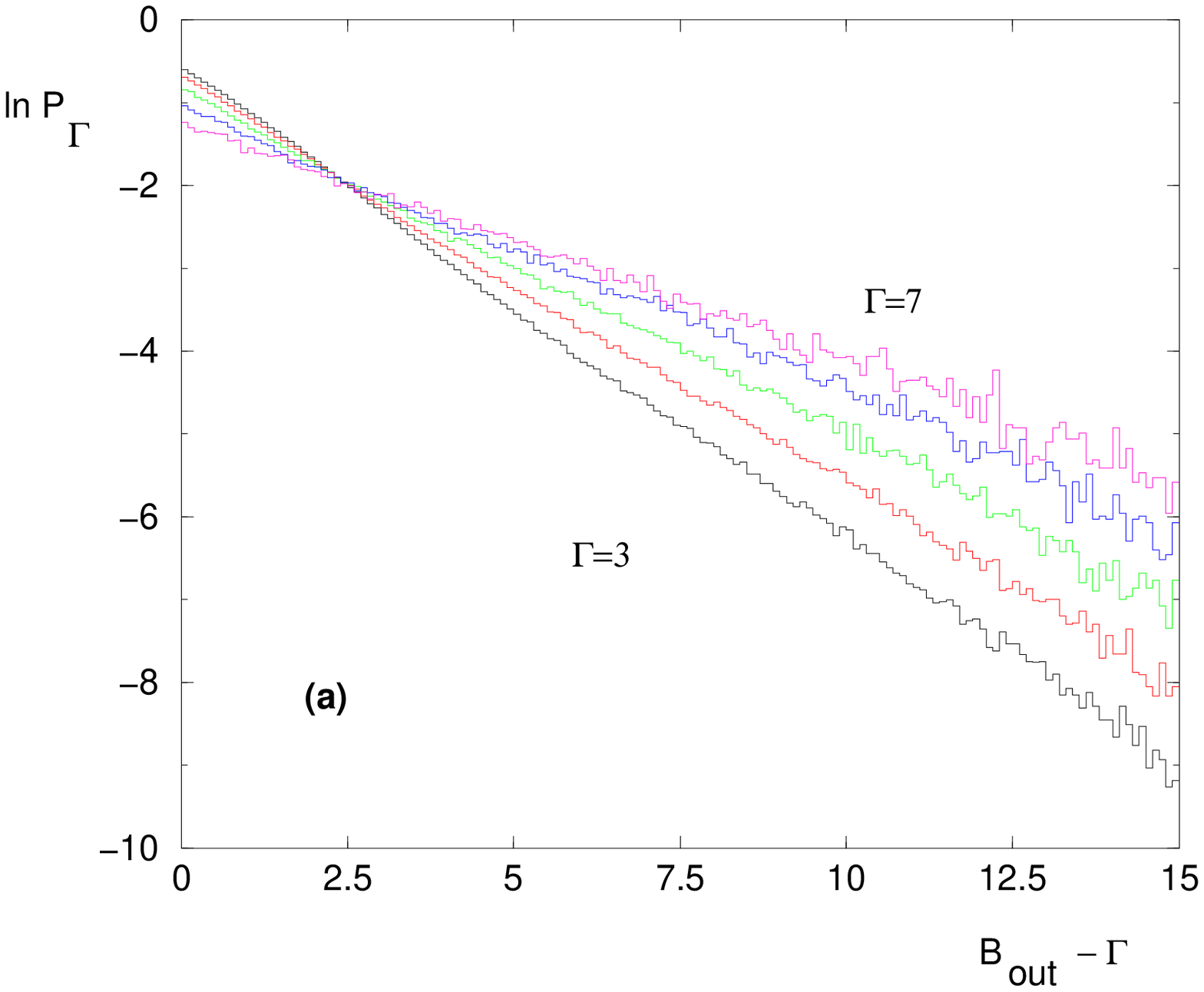}
\hspace{1cm}
\includegraphics[height=6cm]{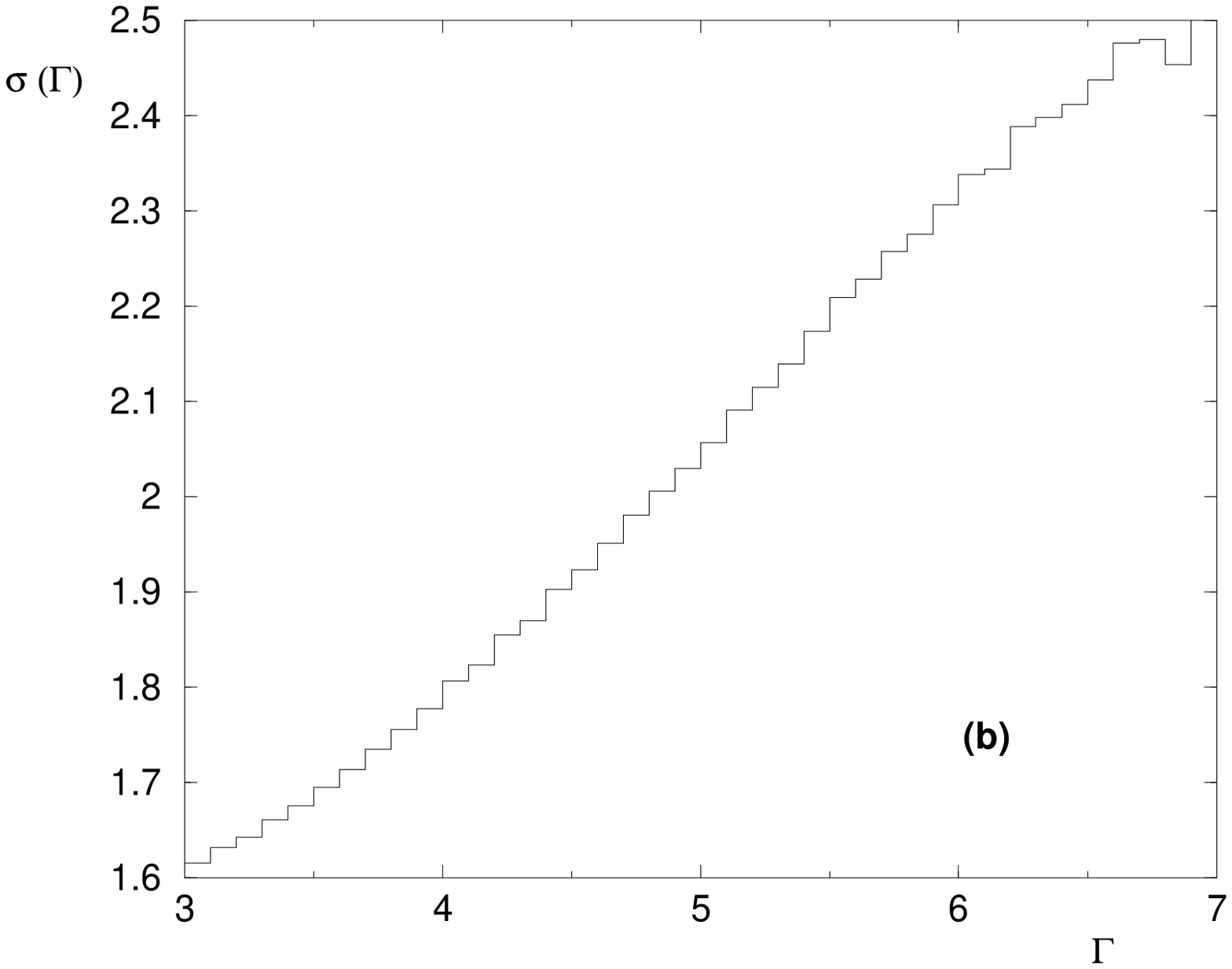}
\caption{ (Color online)
Dynamics of the directed polymer in a two-dimensional random medium :
numerical evidence for the convergence towards an infinite disorder
fixed point (data obtained from the numerical application of
 the full RG rules to $n_s=15.10^4$ disordered samples
for a polymer of length $L=9$ with
  $2^{9}=512$ initial configurations).
(a) 
Flow of the probability distribution $P_{\Gamma}(B_{out}-\Gamma)$
of the renormalized exit barriers (see Eq. \ref{pgammabout})
as the RG scale grows $\Gamma=3,4,5,6,7$ :
these distributions follow the exponential form (see Eq. \ref{pexp})
with a scale-dependent width $\sigma(\Gamma)$.
(b) The corresponding 
width $\sigma(\Gamma)$ grows linearly with the RG scale $\Gamma$    }
\label{fighistobarrier}
\end{figure}

As explained above in section \ref{strongdisorderfixedpoint},
the first important observable to consider is the 
distribution of renormalized exit barriers of Eq. \ref{pgammabout}
to see if the width 
$\sigma(\Gamma)$ grows indefinitely with $\Gamma$ (Eq \ref{infinite}).
If this is the case, then the flow is towards some ``infinite disorder
fixed point'' and 
the strong disorder renormalization approach 
becomes asymptotically exact at large time scales.

We show on Fig. \ref{fighistobarrier} our numerical results
for a directed polymer of length $L=9$ (corresponding
to $2^{9}=512$ initial configurations ).
We find that the rescaled distribution of Eq. \ref{pgammabout} is 
very close to the exponential form
(see Fig. \ref{fighistobarrier} a)
\begin{eqnarray}
{\tilde P} (x) \simeq e^{-x}
\label{pexp}
\end{eqnarray}
and that the width $\sigma(\Gamma)$
grows linearly with the RG scale $\Gamma$
\begin{eqnarray}
\sigma(\Gamma) \sim \Gamma
\label{sigmalinear2}
\end{eqnarray}
after an initial transient regime for smaller $\Gamma$
and a final finite-size saturation regime at larger $\Gamma$.
Note that the two properties of Eqs \ref{pexp} and \ref{sigmalinear2}
 seem extremely robust within strong disorder RG
since they hold exactly in soluble models in $d=1$  \cite{review}
and have been also found numerically in quantum models 
in dimension $d>1$ \cite{motrunich}.

\begin{figure}[htbp]
\includegraphics[height=6cm]{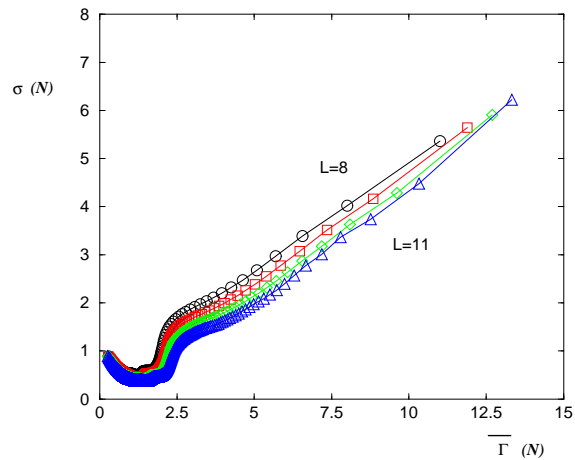}
\hspace{1cm}
\caption{ (Color online)
Numerical evidence of convergence towards an infinite disorder
with data averaged over samples having a fixed
number ${ \cal N}$ of renormalized configurations :
growth of the width $\sigma({ \cal N})$ of the renormalized exit barrier 
with respect to the averaged minimal exit barrier 
$\overline{\Gamma}({ \cal N})$
existing in the system. Note that the linear growth characterizing
 the infinite disorder fixed point
only appears after an initial transient regime.   }
\label{fighistobarriernga}
\end{figure}

The results shown on Fig. \ref{fighistobarrier} have been obtain
by averaging ``at fixed RG scale $\Gamma$'',
i.e. by collecting over many samples 
histograms of renormalized exit barriers when
the last decimated exit barrier  is
within a window of width $\Delta \Gamma = 0.1$ around $\Gamma$
(see section \ref{fixedgamma} for more details).
However as explained above in section \ref{fixednconfig},
it is also interesting to make averages 
over disordered samples having the same number ${ \cal N}$ of surviving 
renormalized configurations. 
For each ${ \cal N}$, we have measured 
the average $\overline{\Gamma}({ \cal N})$
of the minimal exit barrier remaining in the system
and the width $\sigma ({ \cal N})$ of all the remaining exit barriers.
The parametric plot of $\sigma ({ \cal N})$ as a function of
 $\overline{\Gamma}({ \cal N})$ is shown on 
Fig \ref{fighistobarriernga} for the four sizes $L=8,9,10,11$ :
the linear growth characterizing the infinite disorder fixed point
only appears after an initial transient regime, as 
already noted in numerical studies of strong disorder RG
of quantum models in dimension $d>1$ \cite{motrunich}.

\subsection{  Growth of the coherence length $l_{\Gamma}$  }

As explained above, it is convenient to define the coherence length
$l_{\Gamma}$ from 
the number ${ \cal N}_{\Gamma}$ of surviving 
configurations via Eq. \ref{ngamma}.
The barrier exponent $\psi$ of Eqs \ref{defpsi} and 
\ref{typltime} of the Introduction
is the exponent governing the growth of the coherence length at large scale
\begin{eqnarray}
l_{\Gamma} \opsimeq_{\Gamma \to \infty} c \Gamma^{\frac{1}{\psi }} 
\label{defpsi2}
\end{eqnarray}

\begin{figure}[htbp]
\includegraphics[height=6cm]{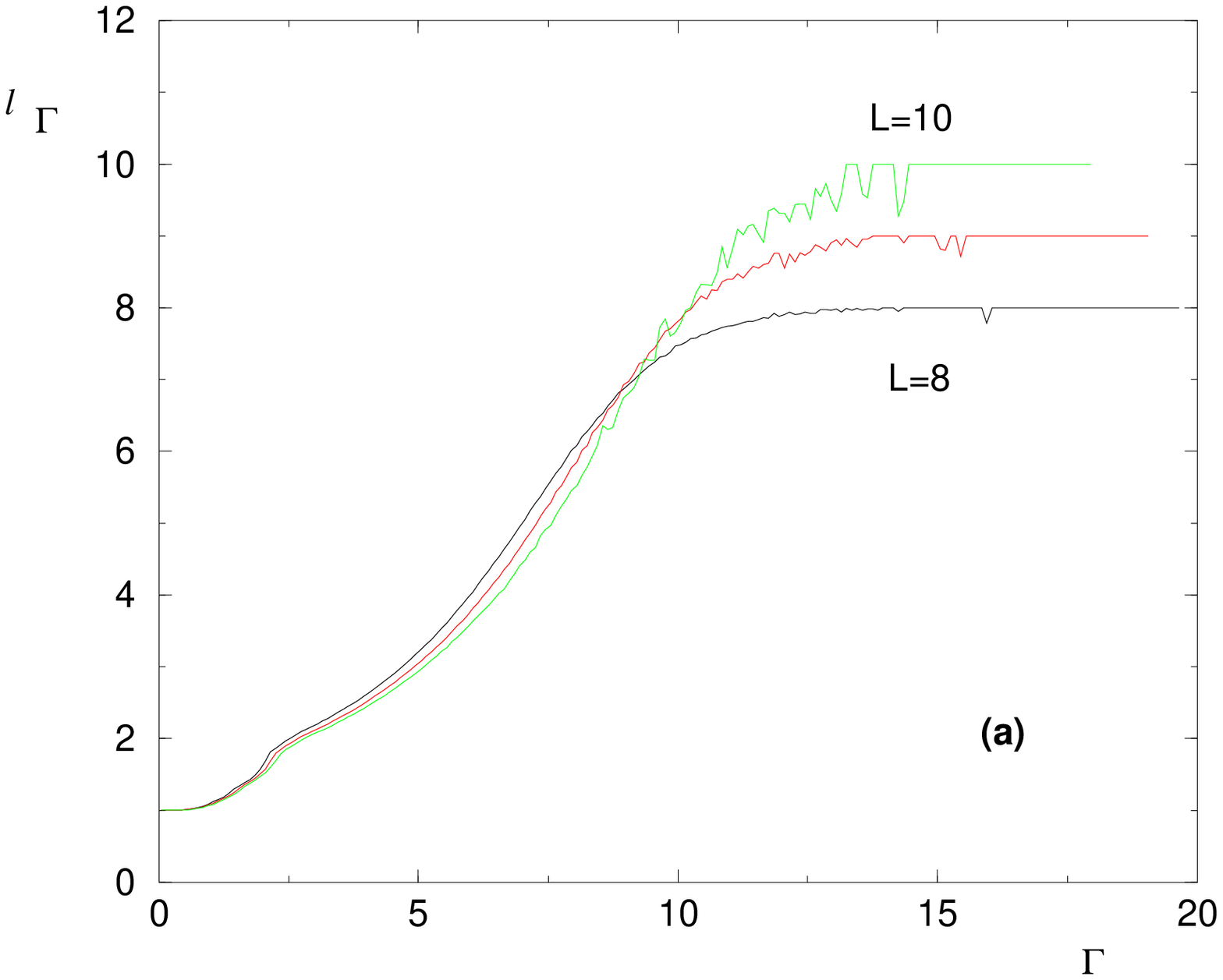}
\hspace{1cm}
\includegraphics[height=6cm]{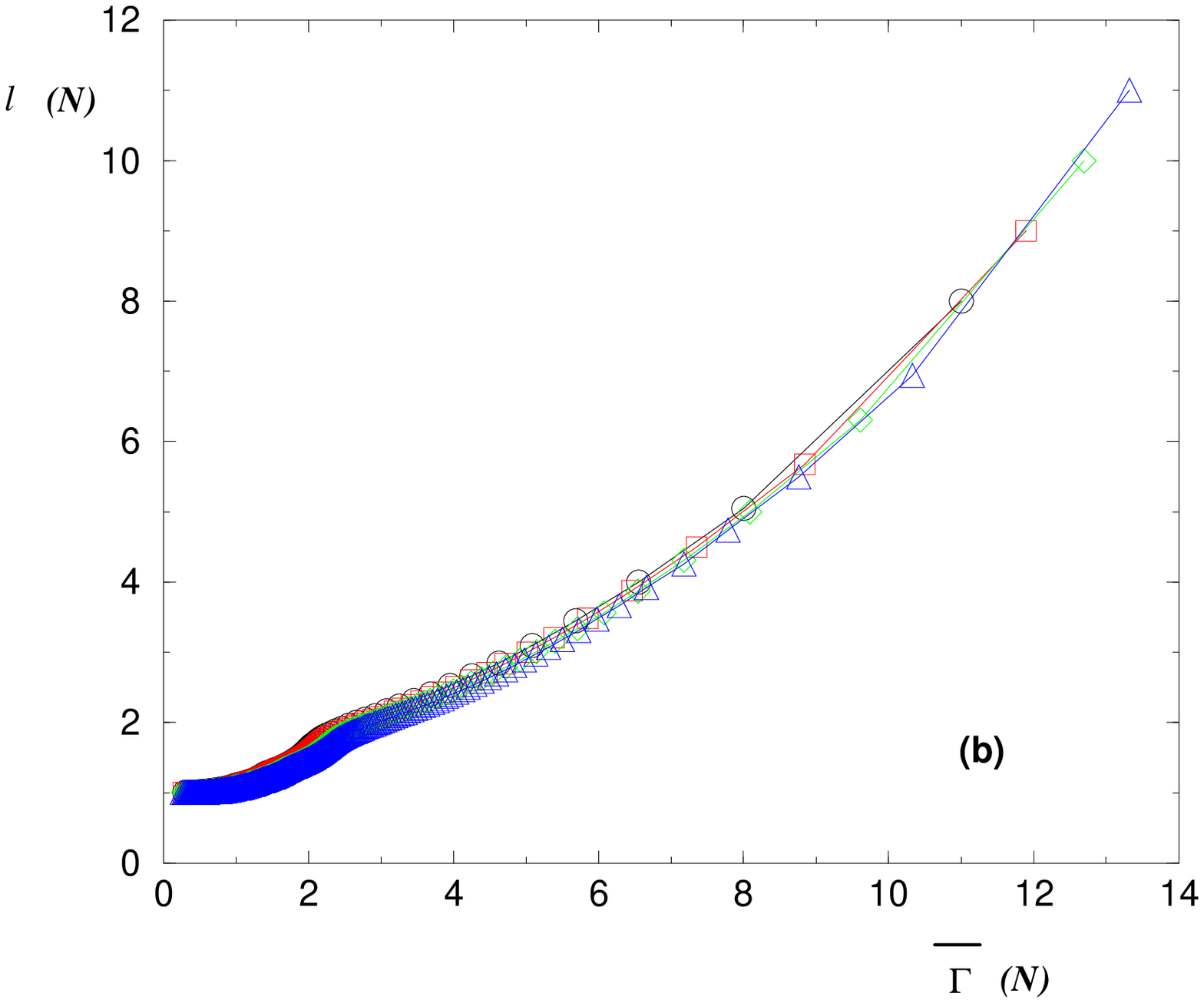}
\caption{ (Color online) Growth of the coherence length $l$ 
with the RG scale $\Gamma$ of the renormalized barriers
(a) Data obtained at fixed RG scale $\Gamma$  for the sizes $L=8,9,10$ :
the coherence length $l_{\Gamma}$ is obtained from the number 
${ \cal N}_{\Gamma} $ of
surviving configurations measured at RG scale $\Gamma$ via Eq. \ref{ngamma}.
(b) Data obtained at a fixed number ${ \cal N}$ of surviving configurations
(i.e. at a fixed coherence length $l$),
for a polymer of length $L=8,9,10,11$.
The horizontal axis $\overline{\Gamma}({ \cal N} )$ represents  the average
of the minimal exit barrier remaining in the system.
 }
\label{figlgamma}
\end{figure}

We show on Fig. \ref{figlgamma} our numerical results concerning
the relation between the barrier scale and the length scale,
within the two ensembles already introduced :

(a) The data corresponding to a fixed RG scale $\Gamma$ 
(see section \ref{fixedgamma}) are shown on Fig. \ref{figlgamma} (a).
The growth of the coherence length $l_{\Gamma}$
as a function of the RG scale $\Gamma$ is shown for the three sizes
$L=8,9,10$ ($L=11$ is not shown here because the data are too noisy) :
after a common growth, the curves separates because they saturate
by construction at the value $l_{eq}=L$.

(b) The data corresponding to a fixed number of 
${ \cal N}$ of surviving configurations, i.e.
to a fixed coherence length $l$ (via Eq. \ref{ngamma})
are shown on  Fig. \ref{figlgamma} (b).
The horizontal axis then corresponds
to the average $\overline{\Gamma}({ \cal N} )$
of the last decimated exit barrier.

The comparison of (a) and (b) show that, for the coherence length,
 the numerical data obtained in the 
``fixed ${ \cal N}$ ensemble'' display less finite-size effects than
data obtained in ``fixed $\Gamma$ ensemble'' and are thus easier to analyze.

\subsection{ Statistics of the equilibrium time
 of finite systems }

\begin{figure}[htbp]
\includegraphics[height=6cm]{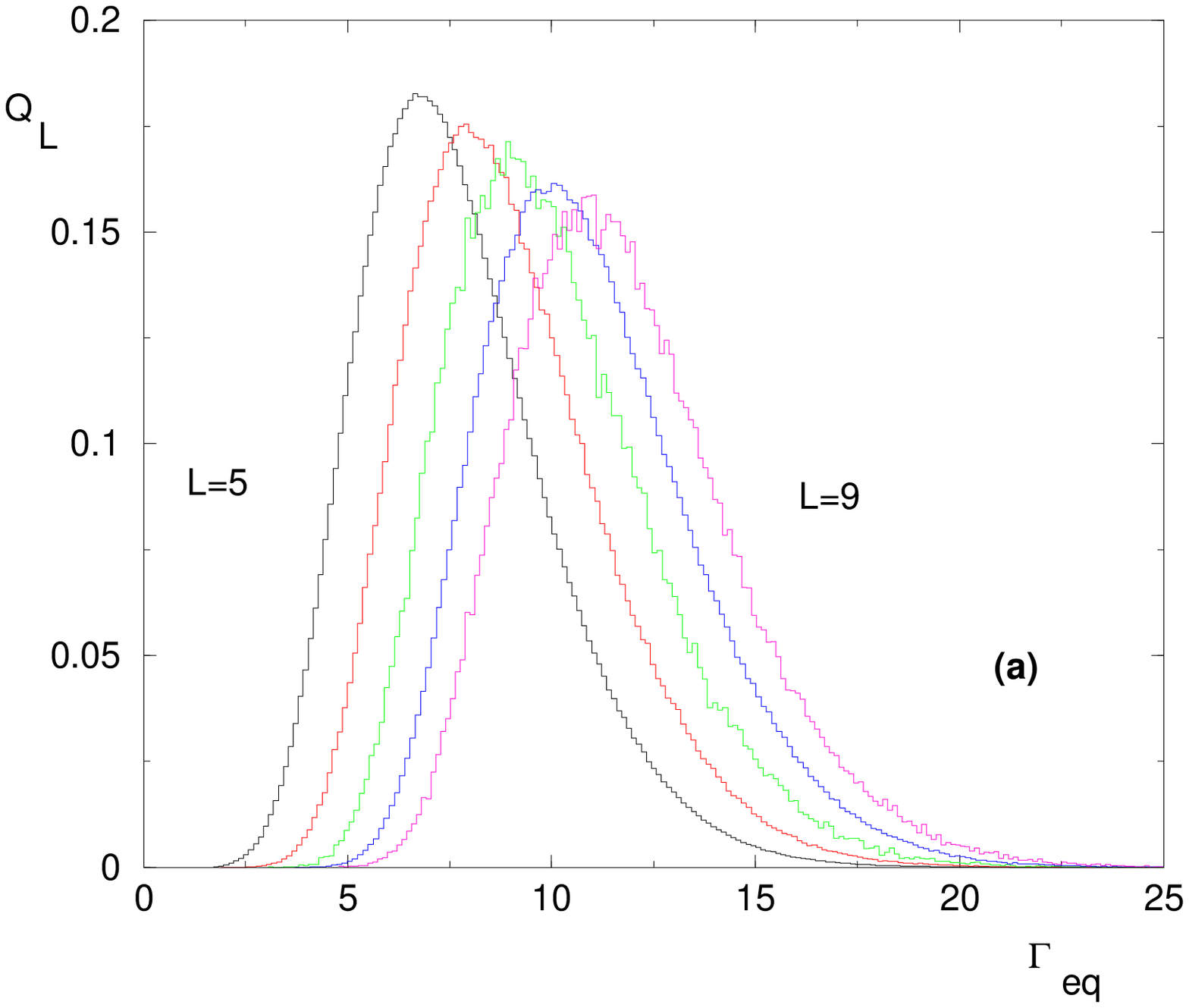}
\hspace{1cm}
\includegraphics[height=6cm]{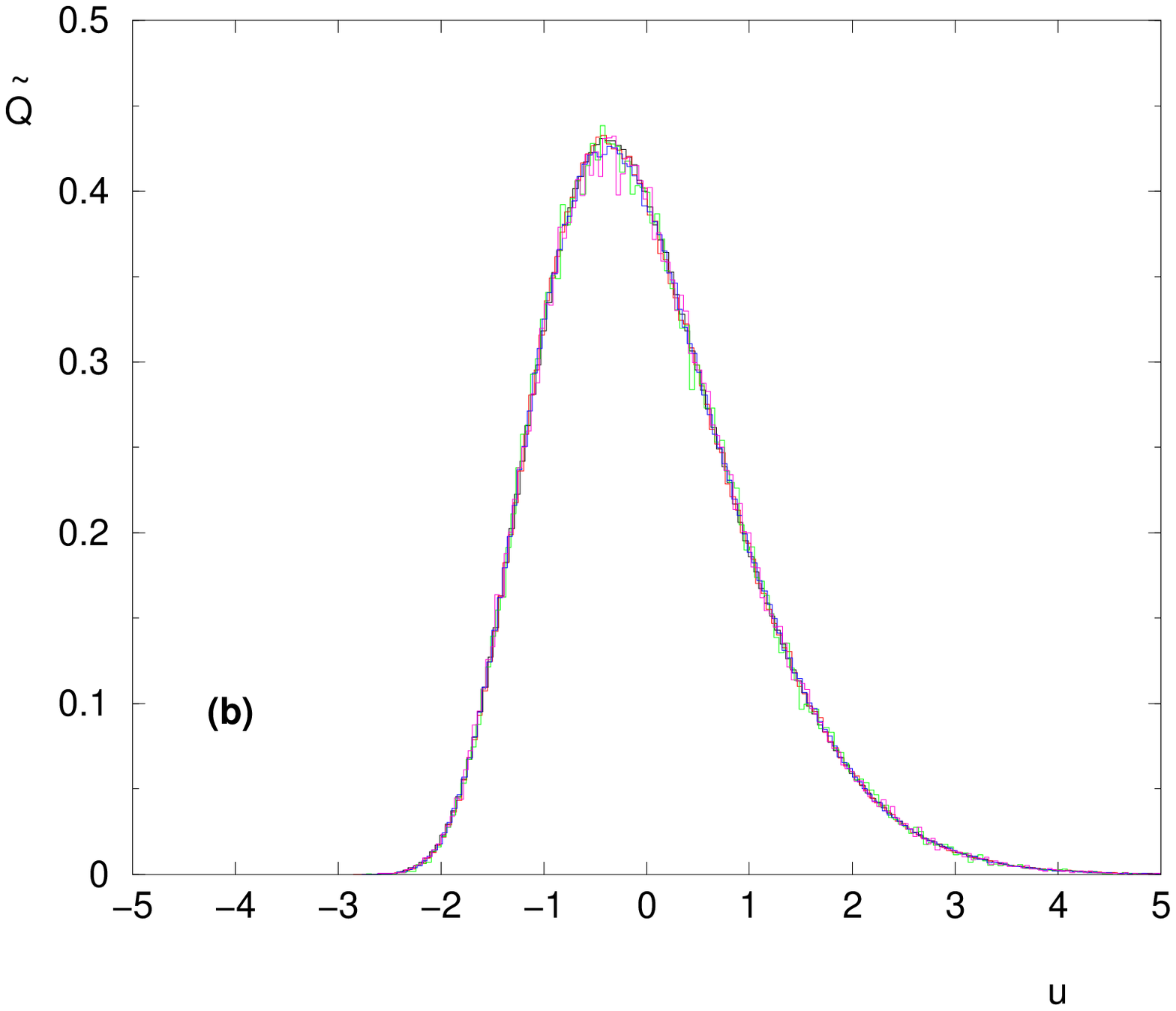}
\caption{ (Color online)
Statistics of the equilibrium time $t_{eq}$ over the disordered samples
for a directed polymer of length $L=5,6,7,8,9 $
 in a two-dimensional random medium   :
(a) Histograms $Q_{L}(\Gamma_{eq}=\ln t_{eq})$
 of the last decimated renormalized exit barrier $ \Gamma_{eq}=\ln t_{eq}$.
(b) Same data in rescaled variables to obtain the 
 rescaled distribution ${\tilde Q}(u)$ of Eq. \ref{qlteq}.
 }
\label{figteq}
\end{figure}

Within the strong disorder RG procedure, the equilibrium time
$t_{eq}$ of a given disordered sample is determined
by the renormalized exit barrier  
\begin{eqnarray}
\Gamma_{eq}= \ln t_{eq}
\label{defgammaeq}
\end{eqnarray}
corresponding to the last decimation process where the two 
biggest metastable valleys merge into a surviving valley
corresponding to thermal equilibrium of the whole sample.
We have measured its probability distribution 
 $Q_{L}(\Gamma_{eq}=\ln t_{eq})$
over the disordered samples of size $L$ as shown on Fig. \ref{figteq} (a).
 The convergence towards a fixed rescaled distribution 
\begin{eqnarray}
Q_{L}(\Gamma_{eq}) \sim  
  \frac{1}{\Delta(L) } {\tilde Q} 
\left( u \equiv \frac{\Gamma_{eq} - \overline{\Gamma_{eq}}(L) }{\Delta(L) }
 \right)
\label{qlteq}
\end{eqnarray}
 is rapid as shown on Fig. \ref{figteq} (b).
However the sizes studied are not sufficient to obtain
a reliable measure of the barrier exponent $\psi$
 the average $\overline{\Gamma_{eq}(L)} \sim L^{\psi}$.
This is why we introduce in the next section simplified RG rules
that are valid near infinite disorder fixed points
and that allow to study numerically bigger system sizes.

\section{ Simplified RG rules near infinite disorder fixed points} 

\label{simpliRG}

\subsection{ Dominance of the preferred exit channel }

Whenever the flow is towards some `infinite-disorder' fixed point,
where the distribution of renormalized exit barriers becomes
 broader and broader
upon iteration (Eqs \ref{pgammabout} and \ref{infinite}),
one expects that the exit rate
out of the decimated configuration ${\cal C}^*$  
\begin{eqnarray}
W_{out} \left( {\cal C}^* \right)  =
 \sum_{i=1}^n  W \left({\cal C}^* \to  {\cal C}_i \right) 
\label{woutcstar}
\end{eqnarray}
will actually be dominated by the preferred exit channel
$i_{pref}$ having the biggest contribution in the sum of Eq. \ref{woutcstar} 
\begin{eqnarray}
W_{out} \left( {\cal C}^* \right)  \simeq
  W \left({\cal C}^* \to  {\cal C}_{i_{pref}} \right) 
\label{wpref}
\end{eqnarray}
i.e. one expects that the probability distribution 
$\pi_{{\cal C}^*} \left({\cal C}_j \right)$ of Eq. \ref{picstar}
will become a delta distribution on the preferred exit channel
up to exponentially small terms 
\begin{eqnarray}
\pi_{{\cal C}^*} \left({\cal C}_j \right) \simeq \delta_{j,i_{pref}}
 +...
\label{picstardelta}
\end{eqnarray}

 The dominance of the preferred exit channel (Eq. \ref{wpref})
near an infinite disorder fixed point will be checked numerically
below for the case of the directed polymer in a two-dimensional
random medium (see section \ref{comparison}). However, we expect that
it holds more generally for the following reasons.
The RG rules with their characteristic multiplicative structure
of Eqs \ref{wijnew} and \ref{picstar}
act directly on the transition rates 
$ W \left({\cal C}_j \to  {\cal C}_i \right) $ between configurations,
whereas the total exit rates $W_{out}$ are derived quantities
obtained by summing over the possible exit channels. The notion
of convergence towards an infinite disorder fixed point
has been defined above by the property that 
the probability distribution of the remaining exit rates $W_{out}$
becomes broader and broader. However we expect that when it happens,
it is because the probability distribution of the individual
transition rates 
$ W \left({\cal C}_j \to  {\cal C}_i \right) $ themselves
 becomes broader and broader, so that the sum in Eq. \ref{woutcstar}
is dominated by the biggest term. A simple one-dimensional example
of this phenomenon is the Sinai model \cite{sinairg},
where each renormalized configuration has always $n=2$ neighbors :
the exit rate to the right and to the left surviving configurations 
follow the infinite disorder scaling form of Eqs \ref{pgammabout} 
and \ref{infinite} : $W_{right} \sim e^{- \Gamma (1+\eta_{right})}$
and $W_{left} \sim e^{- \Gamma (1+\eta_{left})}$ where $\eta_{right}$
and $\eta_{left}$ are independent random variables of order 1 distributed
with the exponential distribution $P(\eta) =e^{-\eta}$. As a consequence,
the exit rate $W_{out} = W_{right}+ W_{left}$ is dominated by
the biggest of the two terms in the limit $\Gamma \to \infty$,
since the probability of degeneracy is of order $1/\Gamma \to 0$.
This type of rare events where the dominance of the preferred exit channel
is not realized will be discussed in more details below in section
\ref{metastable}), but we first state precisely
 how the full RG rules can be simplified when
the preferred exit channel dominates.

\subsection{ Simplified RG rules using the preferred exit channel }

We thus introduce the following simplified RG procedure with respect to the
full RG procedure described in previous section :

(1) the first point is the same ( Eq \ref{defwmax})

(2') among the neighbors $({\cal C}_1,{\cal C}_2,...,{\cal C}_n)$
 of configuration ${\cal C}^*$, find the preferred exit channel $i_{pref}$.
Update the
transitions rates from the $(n-1)$ non-preferred
neighbors $i \ne i_{pref}$ towards $i_{pref}$ by
the approximated rule 
\begin{eqnarray}
W^{new}({\cal C}_i \to {\cal C}_{i_{pref}} ) \simeq
W({\cal C}_i \to {\cal C}_{i_{pref}} )
+ W({\cal C}_i \to {\cal C}^* )
\label{wijnewtowardsipref}
\end{eqnarray}
where the probability distribution $\pi_{{\cal C}^*} \left({\cal C}_j \right)$ of the full RG rule of Eqs \ref{wijnew} and \ref{picstar}
has been replaced by the leading delta function of Eq. \ref{picstardelta}.
Update the transitions rates from $i_{pref}$ towards
the $(n-1)$ non-preferred neighbors $i \ne i_{pref}$ by the full RG rule
of Eqs \ref{wijnew} and \ref{picstar}
\begin{eqnarray}
W^{new}({\cal C}_{i_{pref}}\to {\cal C}_i ) =
W({\cal C}_{i_{pref}} \to {\cal C}_i )
+ W({\cal C}_{i_{pref}} \to {\cal C}^* ) \times 
\frac{W \left({\cal C}^*  \to {\cal C}_i \right)}
{W_{out} \left( {\cal C}^*\right)}
\label{wijnewfromipref}
\end{eqnarray}
Here the full rule is used because the ratio 
$\frac{W \left({\cal C}^*  \to {\cal C}_i \right)}
{W_{out} \left( {\cal C}^*\right)}$ is small and should thus
be evaluated correctly.

In contrast with rule (2), where the increase
 in the total number $N^{rates}$ of transition
rates was only bounded by Eq. \ref{deltanrates},
the rule (2') 
ensures that the total number of renormalized transitions rates
 always decreases
\begin{eqnarray}
\Delta N^{rates} \leq 2 (n-1) -2 n =-2
\label{deltanratespref}
\end{eqnarray}
Moreover, in contrast with rule (2), where 
 the increase in the coordination
number of the $n$ neighbors ${\cal C}_i$ 
was only bounded by Eq. \ref{deltaz}, the rule (2') 
ensures that the coordination numbers
of the non-preferred neighbors do not grow
\begin{eqnarray}
\Delta z^{i \ne i_{pref}} \leq 0
\label{deltaznonpref}
\end{eqnarray}
 and it is only the coordination
number of the preferred neighbor that may grow up to 
\begin{eqnarray}
\Delta z^{i_{pref}} \leq (n-1)-1=n-2 
\label{deltazpref}
\end{eqnarray}

(3') With the rule of Eq. \ref{wijnewtowardsipref}, the exit rates out of
the $(n-1)$ non-preferred neighbors $i \ne i_{pref}$ do not have to be updated
since the exit rate towards 
 ${\cal C}^*$ has been completely transfered to $i_{pref}$.
So the only update of exit rate is for the preferred neighbor $i_{pref}$
via the definition of Eq. \ref{wioutnewactualisation}
or with the equivalent rule of Eq. \ref{wioutnew}.

(4) return to (1)

It is thus clear that these simplified RG rules 
correspond to a substantial gain
from a computational point of view and will allow
to study bigger system sizes.
We will describe in section \ref{numesimpli} 
the numerical results that can be obtained for the directed polymer,
and compare them with the numerical results concerning the full RG rules.
However besides this numerical gain, these simplified
rules have also important implications from a theoretical point of view 
as we now explain.

\subsection{ Interpretation in terms of quasi-equilibrium 
within metastable states} 

\label{metastable}

In the studies on slowly relaxing systems such as disordered systems,
glasses or granular media, it is usual to separate
the dynamics into two parts : there are `fast' degrees of freedom
which rapidly reach local quasi-equilibrium plus a slow non-equilibrium part.
Within the present strong disorder renormalization in configuration space,
these ideas can be applied directly as follows.
To each time $t$, one may associate a set of metastable states
which are labelled by the surviving configurations at the
RG scale $\Gamma= \ln t$. Within each metastable state, configurations are 
quasi-equilibrated, whereas configurations belonging to different
metastable states are still out of equilibrium. 
The slow non-equilibrium part of the dynamics corresponds to the 
evolution of the renormalized valleys with the RG scale :
some valleys disappear and are absorbed by a neighboring valley.

Since at large scale, the RG flows towards an ``infinite disorder'' fixed 
point, the different time scales are effectively very well separated.
As a consequence, we may write, 
as in the Sinai model \cite{sinaimetastablestates}, that
the probability $P({\cal C} t|{\cal C}_0 0)$
 to be in configuration ${\cal C}$
at time $t$ when starting in configuration ${\cal C}_0$
at time $t=0$ is very well approximated by
\begin{eqnarray}
P({\cal C} t|{\cal C}_0 0) \simeq \sum_{V_\Gamma} \frac{1}{Z_{V_\Gamma}}
e^{- \beta E({\cal C})} \theta_{V_\Gamma}({\cal C})
\theta_{V_\Gamma}({\cal C}_0)
\label{valleysum}
\end{eqnarray}
where the sum is over all the renormalized valleys $V_\Gamma$
that are present in the system at the renormalization
scale $\Gamma= \ln t$, and where 
$\theta_{V}({\cal C})$ is the characteristic function 
of the valley $V$, i.e $\theta_{V}({\cal C})=1$ if ${\cal C}$ belongs to the
valley and $\theta_{V}({\cal C})=0$ otherwise. 
The denominator represents the Boltzmann partition function
over the valley $V_\Gamma$
\begin{eqnarray}
Z_{V_\Gamma} = \sum_{{\cal C} \in V_\Gamma}  e^{- \beta E({\cal C})}
 \end{eqnarray}
As discussed in detail in \cite{sinaimetastablestates}, 
the approximation of Eq. \ref{valleysum} breaks down only for
rare events at large times near the infinite disorder fixed point.
More precisely, 
the most important rare events
 that leads to temporary out-of-equilibrium situations
for the set of thermal trajectories starting in the same configuration
${\cal C}_0$ correspond
to the following cases

(i) when the valley $V_{\Gamma}$ containing ${\cal C}_0$ is 
being decimated precisely at an RG scale of order $\Gamma= \ln t$ :
then the thermal packet is broken into two sub-packets, one has already
jumped over the barrier, whereas the other has not jumped yet.
Near the infinite disorder fixed point described by Eq. \ref{pgammabout}
and \ref{sigmalinear2} for the distribution of renormalized exit
barrier, these events occur with a vanishing probability of order
$1/\Gamma=1/(\ln t)$ at large times.

(ii) when the decimation of a valley corresponds to an accidental degeneracy
between the second preferred exit channel and the first preferred exit channel.
Then the thermal packet is also broken into two sub-packets,
one having jumped into the first preferred exit channel
and the other having jumped into the second preferred exit channel.
Again, near the infinite disorder fixed point 
 these events occur with a vanishing probability of order
$1/\Gamma=1/(\ln t)$ at large times.

This discussion shows that the asymptotic
dominance of the preferred exit channel near the infinite disorder
fixed point is actually crucial to obtain quasi-equilibrium
within the visited region of phase space at a given large time $t$.
In particular, if the degeneracy between the second preferred exit channel 
and the first preferred exit channel could occur with a finite probability,
then finite contributions of out-of-equilibrium situations
at all scales would ruin the quasi-equilibrium approximation of
Eq. \ref{valleysum} : the probability to be in a configuration ${\cal C}$
at time $t$ would not depend only on its energy $E({\cal C})$ 
and on the partition function $Z_{V_{\Gamma}}$ of the renormalized valley
it belongs, but would be instead a very complicated function
of all possible paths from ${\cal C}_0$ to ${\cal C}$ 
with their appropriate dynamical weights.
To better understand the importance of this discussion,
it is useful to recall here a well-identified
 exception of the quasi-equilibrium idea, namely
  the symmetric Bouchaud's trap model in one dimension,
where even in the limit of arbitrary low temperature,
the diffusion front in each sample consists
 in two delta peaks, which are completely out of equilibrium with each other
\cite{traprg} : the weights of these two delta peaks
do not depend on their energies, but instead 
 on the distances to the origin
that determine the probability to reach one before
the other (see \cite{traprg} for more details).
In this trap model, the reason is clear : whenever the particle
escapes from a trap, it jumps either to the right or to the left
with equal probabilities $(1/2)$, i.e. the two possible exit channels
are degenerate by the very definition of the model that imposes
this symmetry. In other disordered models where this degeneracy is not
imposed by a symmetry of the model, this degeneracy can only 
occur accidentally with some probability. The question is then whether
this probability of accidental degeneracy between
the two preferred exit channels remains finite or 
becomes rare (i.e. decays to zero) at large times. 
Within the present strong disorder RG where the flow is towards some
infinite disorder fixed point, the dominance of the preferred exit channel
precisely means that the probability of these accidental degeneracy
decays to zero, so that the quasi-equilibrium approximation of
Eq. \ref{valleysum} becomes asymptotically exact at large times.

\section{ Numerical studies using simplified RG rules}

\label{numesimpli}

\subsection{ Numerical gain with respect to the full RG rules   }

As explained in section \ref{proliferation}, the numerical application
of the full RG rules are limited to small sizes because
the proliferation of neighbors is memory and time consuming.
The simplified RG rules described in section \ref{simpliRG},
that preserve the asymptotic exactness near infinite disorder fixed points,
allow to study much bigger system sizes.
For instance, for the directed polymer in a two dimensional model
introduced previously (section \ref{DPmodel}),
the linear sizes $L$ and the corresponding numbers $n_s(L)$ 
of disordered samples that we have been able to study 
{\it via simplified rules } are 
\begin{eqnarray}
 L && =  7, 8, 9, 10, 11, 12, 13, 14, 15, 16, 17, 18  \nonumber \\
n_s(L) && = 7.10^8, 3.10^8, 10^8, 3.10^7, 10^7, 3.10^6, 7.10^5,18.10^4,4.10^4,   8.10^3 , 1500 , 600
\label{sizessimpliRG}
\end{eqnarray}
that should be compared to the sizes given in Eq. \ref{sizesfullRG}
{\it for the full RG rules}.
In particular, note the difference for the biggest sizes :
the biggest size $L=11$ for the full RG rules corresponds
to an initial number of $2^{11}=2048 $ configurations,
whereas the biggest size $L=18$ for the simplified RG rules corresponds
to an initial number of $2^{18}=262 144 $ configurations.
The numerical gain is thus substantial.

\subsection{ Choice of initial conditions to improve the convergence
towards the fixed point   }

\label{initial}

The simplified RG rules are based on the dominance
of the preferred exit channel, which is realized only
near ''infinite disorder fixed points'', i.e. they will be good
at large RG scales, but not during the first RG steps.
As a consequence, it is important
to stress here the two different aims of the numerical studies
based on full and simplified RG rules respectively

(i) the aim of the {\it full RG rules } is to study
whether the true microscopic model under interest indeed
flows towards an ``infinite disorder fixed point''
where the width of the renormalized barriers distribution
grows without bounds. This is what we have checked in section \ref{numefull}
for the directed polymer, using as initial transition rates
the 'true' Metropolis transition rates of Eq. \ref{metropolis}.

(ii)  the aim of the {\it simplified RG rules } is to study 
directly the properties of the ``infinite disorder fixed point''.
As a consequence here, we do not wish to use as initial transition rates
the 'true' Metropolis transition rates of Eq. \ref{metropolis},
but instead the initial conditions that reduce the transient as much as
possible, i.e. the initial conditions that are the closest 
to the fixed point properties. 
As in the strong disorder RG studies of quantum models  
 where the same strategy was used \cite{motrunich}, 
one would like to choose an initial condition where the 
probability distribution of barriers is already exponential,
i.e. of the same form observed at large scale (see Eq. \ref{pexp})
when applying the full RG rules to the Metropolis initial condition.
For the spatial properties however, since the random correlated structure
generated by the RG flow is difficult to characterize,
one is restricted as in \cite{motrunich} to start from the regular
lattice structure of the microscopic model.

We now describe more precisely the initial conditions
we have used for our numerical studies for the directed polymer.
Instead of the Gaussian energies of Eq. \ref{epsgauss},
we have drawn sites energies from the exponential distribution
\begin{eqnarray}
\rho (\epsilon ) =\theta(\epsilon \leq 0) \  e^{\epsilon}
\label{rhoexp}
\end{eqnarray}
Then to choose the transition rates, there is still some freedom
within the detailed balance condition,
since Eq. \ref{detailedbalance} simply means that
\begin{eqnarray}
 e^{- E({\cal C}) /T} W \left( {\cal C} \to {\cal C}'  \right)
= e^{- E({\cal C}') /T} W \left( {\cal C}' \to {\cal C}  \right)
= \Delta({\cal C},{\cal C}')
\label{detailedbalance2}
\end{eqnarray}
where $\Delta({\cal C},{\cal C}')=\Delta({\cal C}',{\cal C})$
represents some arbitrary symmetric barrier. 
Since the dominance of the preferred exit channel
has a meaning only if all directions are 'ascending',
we have chosen to avoid the presence of 'descending' directions
 in the initial condition. Since the site energies of Eq. \ref{rhoexp}
are all negative, the energies of the configurations inherit the same
property, and we have thus chosen the following form for the 
symmetric barrier
\begin{eqnarray}
\Delta({\cal C},{\cal C}') 
= e^{ E({\cal C}) /T +E({\cal C}') /T} 
\label{choixdeltasymbarrier}
\end{eqnarray}
This corresponds to the following transition rates
\begin{eqnarray}
W \left( {\cal C} \to {\cal C}'  \right)
= e^{2 E({\cal C}) /T +E({\cal C}') /T} 
\label{choixwini}
\end{eqnarray}
i.e. the barriers of the initial condition are all positive
and given by
\begin{eqnarray}
B \left( {\cal C} \to {\cal C}'  \right)
= - 2 E({\cal C}) /T - E({\cal C}') /T 
\label{choixbini}
\end{eqnarray}

We now describe in the remaining of this section the numerical results
obtained by applying the simplified RG rules starting from
this initial condition.

\begin{figure}[htbp]
\includegraphics[height=6cm]{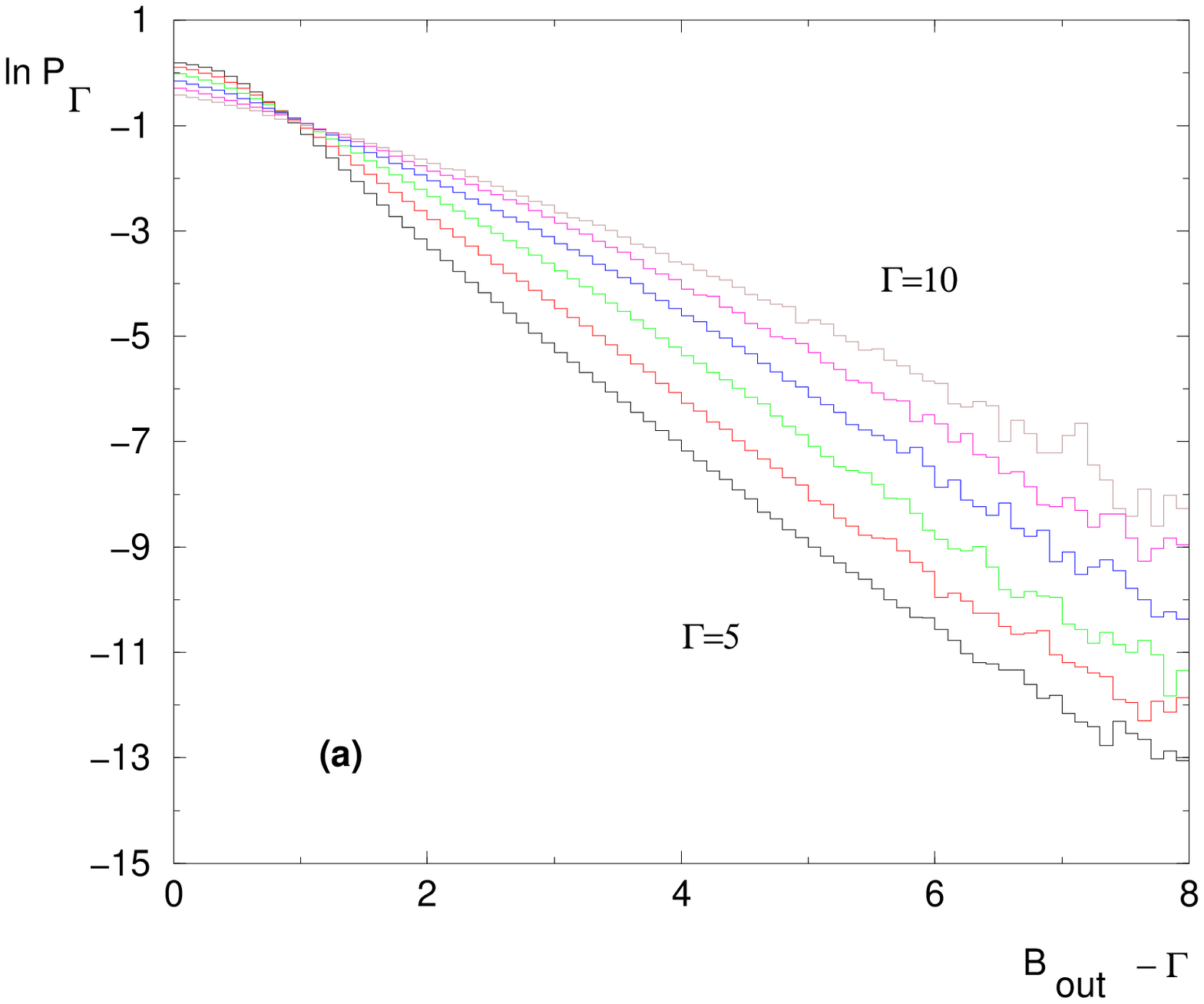}
\hspace{1cm}
\includegraphics[height=6cm]{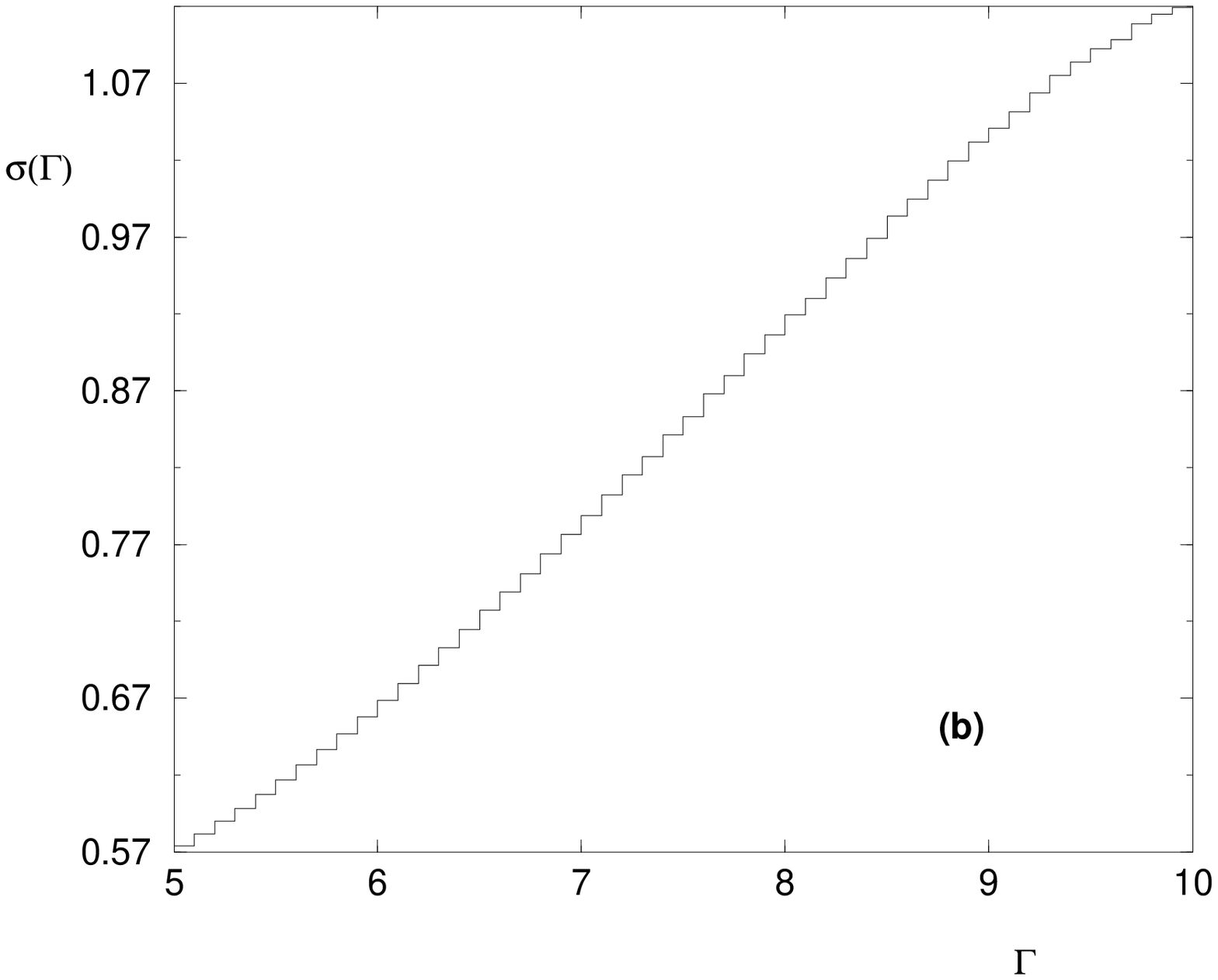}
\caption{ (Color online) Histograms of renormalized exit barrier for
a directed polymer of length $L=15$ ( corresponding to $2^{15}= 32768$
initial configurations) using simplified RG rules
(a) 
Flow of the probability distribution $P_{\Gamma}(B_{out}-\Gamma)$
of the renormalized exit barriers (see Eq. \ref{pgammabout})
as the RG scale grows $\Gamma=5,6,7,8,9,10$ :
(b) Growth of the corresponding 
width $\sigma(\Gamma)$ with the RG scale $\Gamma$.
(These data should be compared with
the equivalent results shown on Fig. \ref{fighistobarrier}
corresponding to the application of full RG rules on a smaller polymer
of length $L=9$)  }
\label{fighistobarriersimpli}
\end{figure}

\subsection{Probability distribution of the renormalized exit barriers   }

We show on Fig. \ref{fighistobarriersimpli} the histograms
of renormalized exit barriers obtained via simplified RG rules
for a polymer of length $L=15$ ( corresponding to $2^{15}= 32768$
initial configurations), 
that should be compared with Fig. \ref{fighistobarrier}
showing the equivalent results obtained via the
full RG rules for a polymer of length $L=9$. 
The important point is the linear growth 
of the width $\sigma(\Gamma) \sim \Gamma$ shown on Fig. 
\ref{fighistobarriersimpli} (b).
The deviations from the exponential distribution
 visible on Fig. \ref{fighistobarriersimpli} (a),
in particular the curvatures near the origin,
indicate that our choice of initial conditions is not optimal :
the problem is that the distribution of the
initial barriers of Eq. \ref{choixbini}
is not exponential near the origin.
However, in the absence of a much better idea for the initial conditions
that would reduce the transients, we have chosen to keep the initial
conditions described in section \ref{initial}.

\subsection{ Growth of the coherence length $l_{\Gamma}$  }

\begin{figure}[htbp]
\includegraphics[height=6cm]{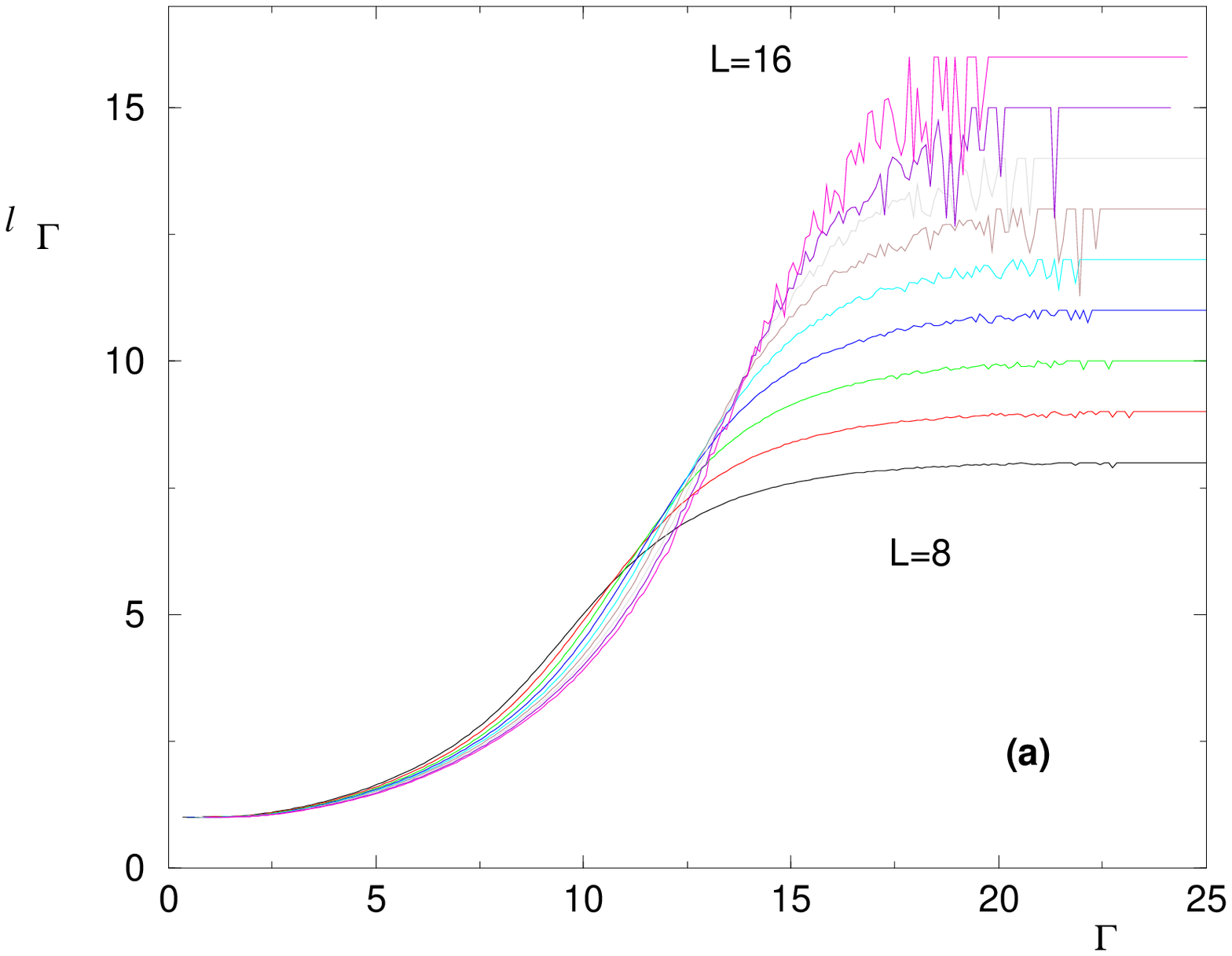}
\hspace{1cm}
 \includegraphics[height=6cm]{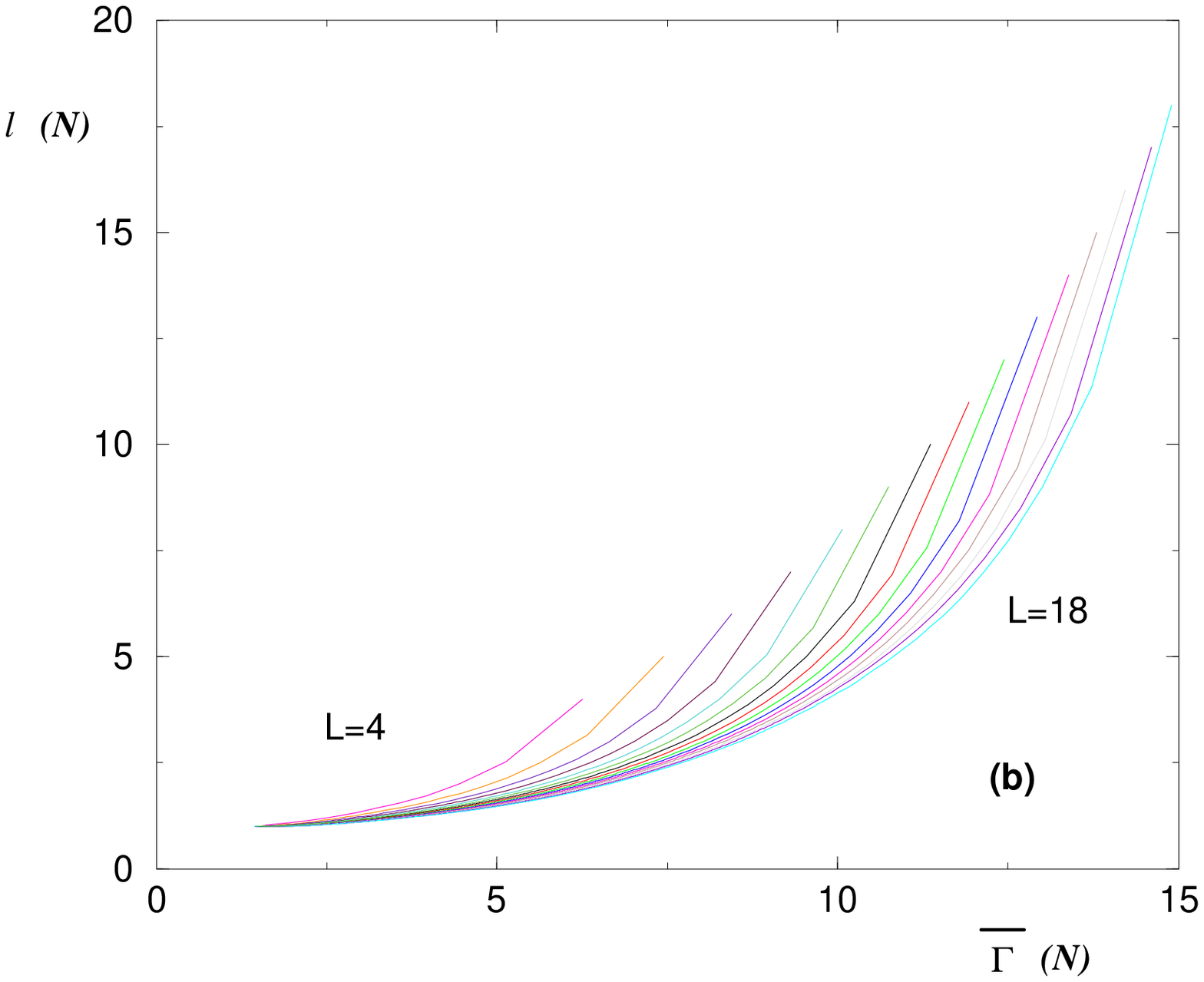}
\caption{ (Color online)
Growth of the coherence length $l_{\Gamma}$ 
with the RG scale $\Gamma$ via simplified RG rules
(a) Data obtained at fixed RG scale $\Gamma$ for the sizes $8 \leq L \leq 16$:
the coherence length $l_{\Gamma}$ is obtained from the number 
${ \cal N}_{\Gamma}$ of
surviving configurations measured at RG scale $\Gamma$ via Eq. \ref{ngamma}
(b) Data obtained at a fixed number ${ \cal N}$ of surviving configurations,
(i.e. at a fixed coherence length $l$), for the sizes $4 \leq  L \leq 18$.
The horizontal axis $\overline{\Gamma}({ \cal N} )$ represents  the average
of the minimal exit barrier remaining in the system.
(These data should be compared with the equivalent results shown
on Fig \ref{figlgamma} obtained
via full RG rules on smaller systems)
 }
\label{figlgammasimpli}
\end{figure}

We show on Fig. \ref{figlgammasimpli} our numerical results concerning
the relation between the barrier scale and the length scale,
within the two ensembles already introduced 
( these data should be compared with 
the equivalent results of Fig. \ref{figlgamma}  for the full RG rules) :

(a) The data corresponding to a fixed RG scale $\Gamma$ 
(see section \ref{fixedgamma}) are shown on Fig. \ref{figlgammasimpli} (a).
The growth of the coherence length $l_{\Gamma}$
as a function of the RG scale $\Gamma$ is shown for the sizes
$8 \leq L \leq 16$ : the curvature is more and more pronounced
before the finite-size saturation at the value $l_{eq}=L$.

(b) The data corresponding to a fixed number of 
${ \cal N}$ of surviving configurations, i.e.
to a fixed coherence length $l$ (via Eq. \ref{ngamma})
are shown on  Fig. \ref{figlgammasimpli} (b).
The horizontal axis then corresponds
to the average $\overline{\Gamma}({ \cal N} )$
of the last decimated exit barrier $\Gamma$.

\subsection{ Statistics of the equilibrium time of finite
systems   }

\begin{figure}[htbp]
\includegraphics[height=6cm]{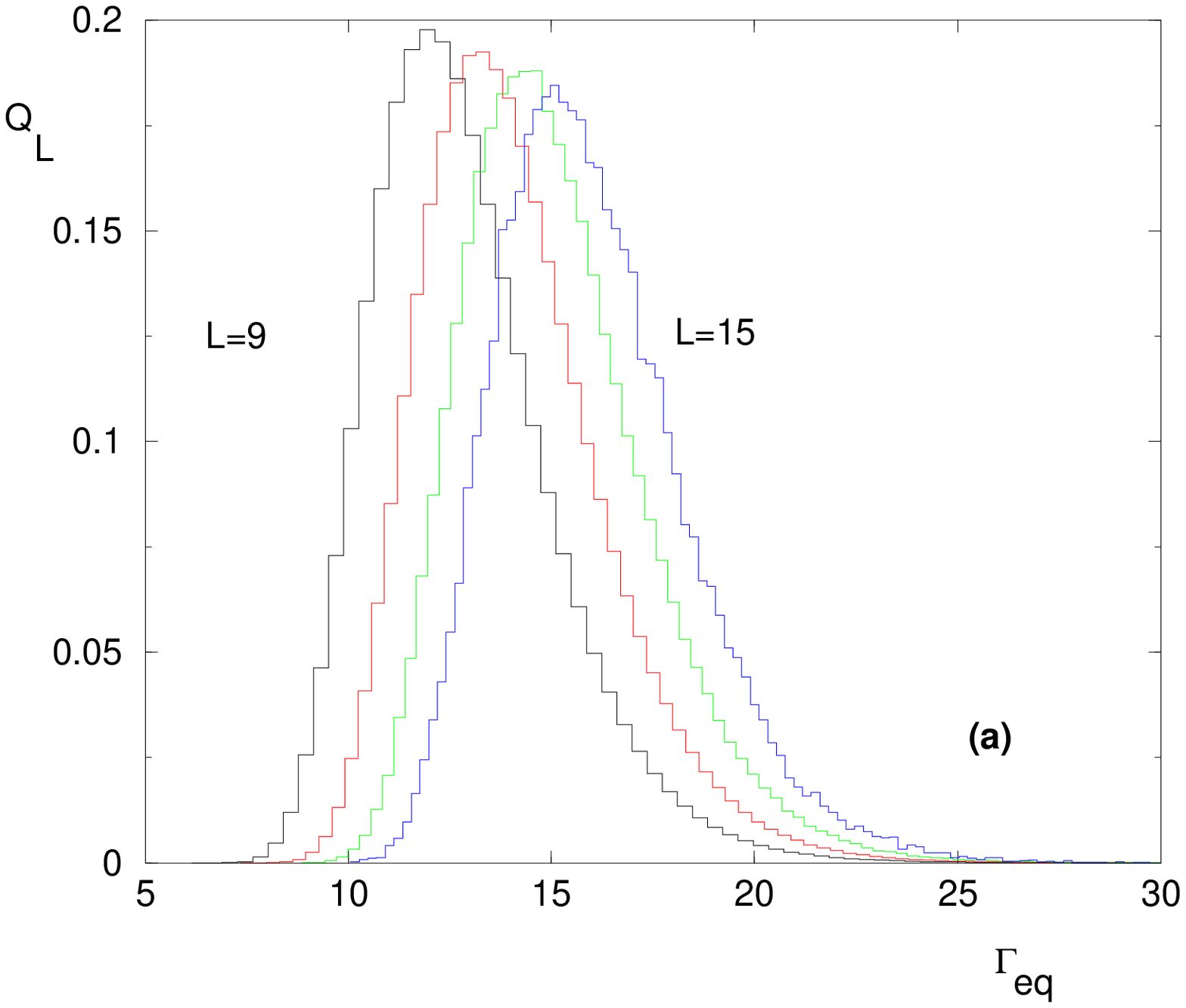}
\hspace{1cm}
\includegraphics[height=6cm]{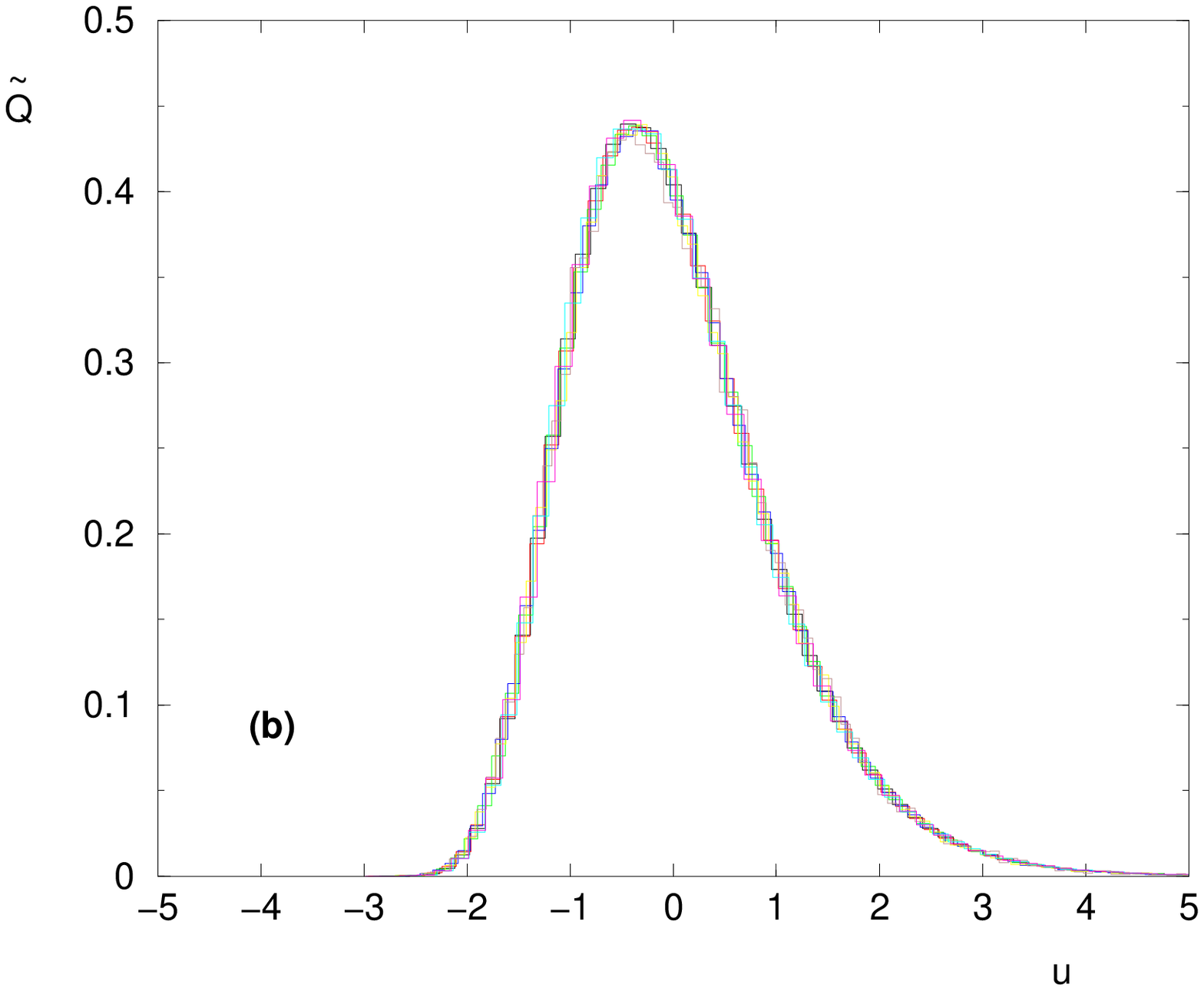}
\caption{ (Color online)
Statistics of the equilibrium time $t_{eq}$ over the disordered samples
of a given length $L$
using simplified RG rules  :
(a) Histograms $Q_{L}(\Gamma_{eq}=\ln t_{eq})$
 of the last decimated renormalized exit barrier $ \Gamma_{eq}=\ln t_{eq}$
for the four sizes $L=9,11,13,15$
 (other sizes have not been shown for clarity).
(b) Rescaled distribution ${\tilde Q}(u)$ (see Eq. \ref{qlteq})
for the sizes $8 \leq L \leq 15$.
(These data should be compared with the equivalent results shown
on Fig \ref{figteq} obtained
via full RG rules on smaller systems) }
\label{figteqsimpli}
\end{figure}

We shown on Fig. \ref{figteqsimpli} the numerical results for
the statistics of the equilibrium time $t_{eq}$ over disordered samples
of a given length $L$ obtained via the simplified RG rules,
that should be compared with the equivalent results of Fig. \ref{figteq}
obtained via the full RG rules.
The rescaled distributions shown on Fig. \ref{figteqsimpli} (b)
are very stable as the size $L$ changes, as on Fig. \ref{figteq} (b)
for the full RG rules.

\subsection{ Comparison of the numerical obtained via full RG rules and via
simplified RG rules   }

\label{comparison}

\begin{figure}[htbp]
\includegraphics[height=6cm]{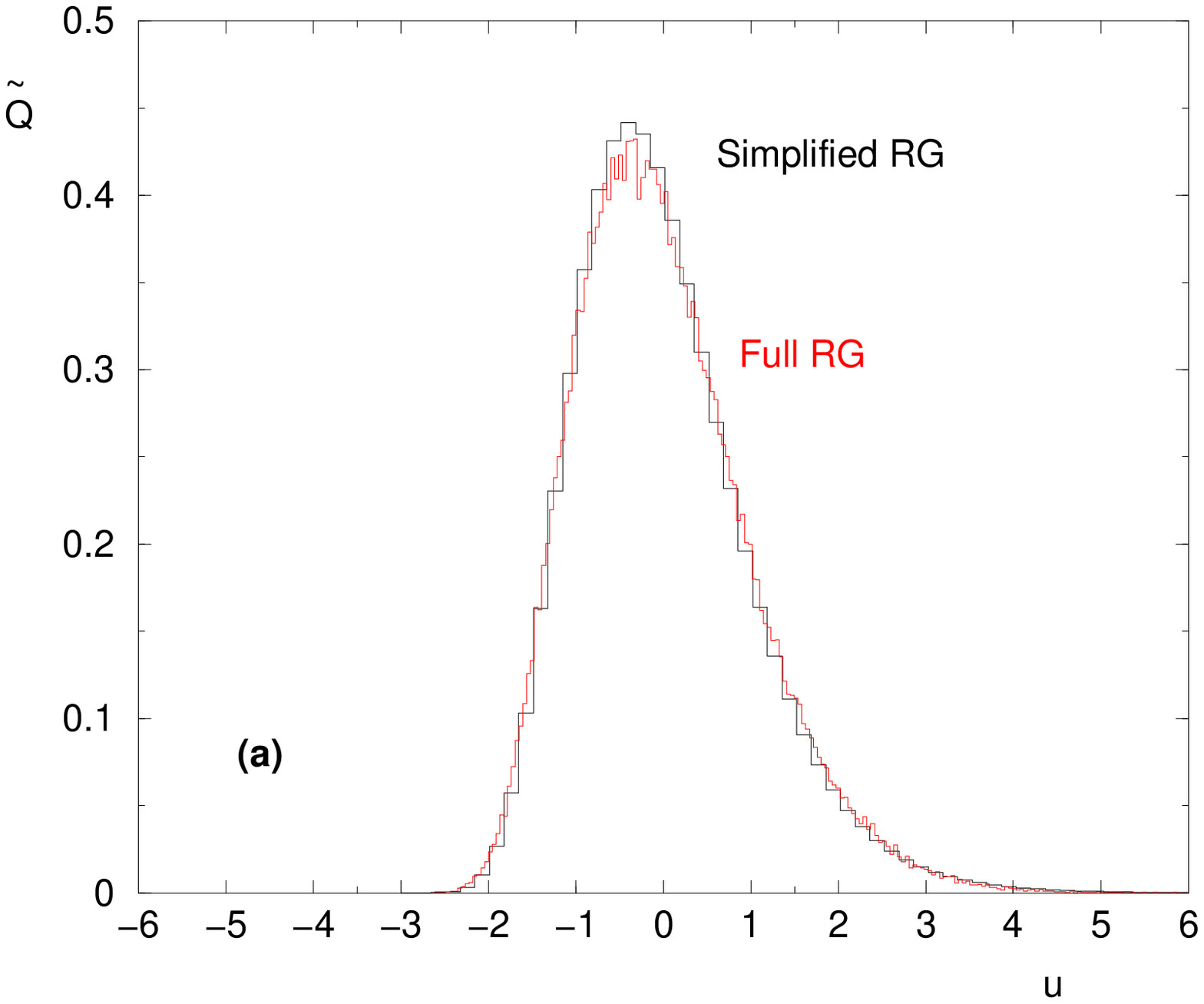}
\hspace{1cm}
\includegraphics[height=6cm]{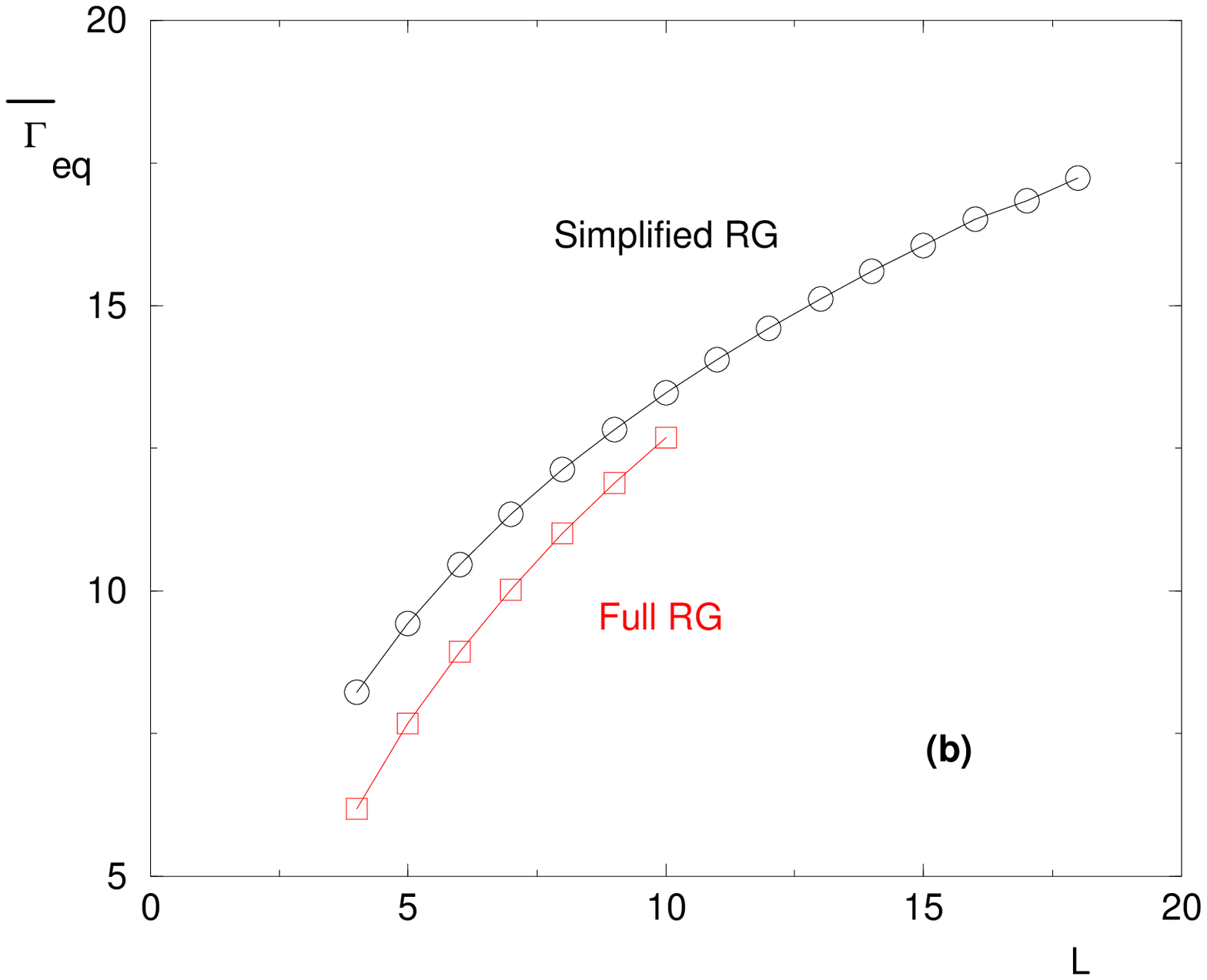}
\caption{ (Color online)
Comparison of data obtained via full RG rules and via simplified RG rules
(a) The rescaled histograms
${\tilde Q}(u)$ (see Eq. \ref{qlteq}) 
obtained via the full RG rules and via the simplified RG rules
respectively, coincide : 
this shows that the simplified RG rules capture correctly
the fixed point properties.
(b) Growth of the averaged equilibrium barrier $\overline{\Gamma_{eq}}(L)$
a function of the system size $ L$ :
the data obtained via full RG rules are limited to the sizes
$4 \leq L \leq 10$, whereas 
the data obtained via simplified RG rules 
are for the sizes
$4 \leq L \leq 18$. The numerical gain is thus substantial.
  }
\label{figteqcompare}
\end{figure}

As explained above in section \ref{initial},
we have used different initial conditions for
our numerical studies of full RG rules and of simplified RG rules,
so we cannot compare the 'numbers' obtained, but we should
compare the stable properties of the fixed point that do not
depend on the microscopic details, i.e. on the details of the
initial condition.
A good test is for instance the rescaled probability distribution
of the equilibrium time $t_{eq}$ over disordered samples
of a given length $L$, that was found to be stable
 with respect to the value of $L$ 
both for the full RG rules (see Fig. \ref{figteq} (b))
and for the simplified  RG rules (see Fig. \ref{figteqsimpli} (b)).
 As shown on Fig. \ref{figteqcompare} (a),
these rescaled probability distributions obtained via the two RG rules
indeed coincide.
This agreement is a strong numerical evidence that 
the simplified RG rules capture correctly
the properties of the fixed point.
This is very important numerically, since the
simplified RG rules allow to study much bigger sizes than the full
RG rules, and we compare for instance on Fig. \ref{figteqcompare} b
the data obtained via the two methods for the
growth of the averaged equilibrium barrier $\overline{\Gamma_{eq}}(L)$
with the length $L$.
Whereas the sizes studied via full RG rules are not sufficient to obtain
a reliable measure of the asymptotic barrier exponent $\psi$,
a direct two-parameter power-law fit
 $\overline{\Gamma_{eq}}(L) =a_0 L^{\psi}$
 of the data obtained via simplified RG rules gives a value of order
\begin{eqnarray}
\psi \sim 0.47 
\label{psivalue}
\end{eqnarray}
This estimate is of course not expected to be very precise, 
as any critical exponent measured in disordered samples
of limited sizes,
but it is is nevertheless rather close
to the best value $\psi \sim 0.49$ presently available
that has obtained by Monte-Carlo simulation
of the Langevin dynamics \cite{rosso}.
Moreover, we have checked that the above value is quite stable
when we analyze the various data on the coherence length 
presented above, either in the ensemble at fixed RG scale $\Gamma$
or in the ensemble at fixed ${\cal N}$.

\subsection{ Debate on the value of the barrier exponent }

An important physical issue is whether the barrier exponent $\psi$
is equal to the droplet exponent $\theta$ of the statics,
 which is exactly known
to be $\theta=1/3$ for the directed polymer in a two dimensional medium.
Although the assumption $\psi=\theta=1/3$ is made systematically
in the literature since the very first paper 
\cite{Hus_Hen} introducing the model, and is sometimes
considered as established up to possible
 logarithmic corrections \cite{drossel}, we have explained elsewhere
 \cite{conjecturepsi} why the equality $\psi=\theta$ is far from obvious
within the droplet scaling theory where the only bounds
are $\theta \leq \psi \leq d-1$ \cite{Fis_Hus},
and where already in the statics,
free-energy fluctuations and energy fluctuations involve
the different exponents $\theta=1/3$ and $d_s/2=1/2$ \cite{Fis_Hus_DP}.
 Moreover in other disordered models like spin-glasses, 
 the barrier exponent $\psi$ is expected to be
strictly bigger than the droplet exponent $\theta$,
 because they are distinct below the lower critical dimension :
in dimension $d=1$, the exact solution \cite{heidelberg} yields
$  \psi_{1d}=0  > \theta_{1d}=-1$, 
and in dimension $d=2$, these two exponents do not have the same sign
$ \psi_{2d} >0 > \theta_{2d} $.
Despite their measure $\psi \simeq 0.49$,
the conclusion of the authors of \cite{rosso}
that believe in the identity $\psi=\theta=1/3$,
is that barriers contain strong logarithmic corrections
$B(L) \sim L^{1/3} (\ln L)^{\mu}$.
Our interpretation is on the contrary that the measured value 
$\psi \sim 0.49$ in  \cite{rosso}
and our present estimate of Eq. \ref{psivalue}
could very well be the correct order of magnitude,
and thus strictly bigger than the droplet exponent $\theta=1/3$.

\section{ Physical interpretation of the barrier exponent $\psi$ }

\label{secpsi}

In this section, we explain how the barrier exponent $\psi$
that relates time and length scales (Eq. \ref{defpsi})
depends on the spatial connectivity of the renormalized degrees of freedom.
In Section \ref{fixednconfig}, we have seen how
to associate to each RG scale $\Gamma$
some coherence length $l_{\Gamma}$, such that the number $n_{\Gamma}$
of renormalized degrees of freedom is
\begin{eqnarray}
 n_{\Gamma} \equiv  \frac{L}{l_{\Gamma}}
\label{ndegrees}
\end{eqnarray}
and the number of ${ \cal N}_{\Gamma}$ of surviving 
configurations at scale $\Gamma$ reads (Eq. \ref{ngamma})
\begin{eqnarray}
 { \cal N}_{\Gamma} \equiv  2^{n_{\Gamma}}
\label{ngamma2}
\end{eqnarray}

On one hand, the decrease of the number of surviving configurations read
\begin{eqnarray}
 d{ \cal N}_{\Gamma} = - { \cal N}_{\Gamma} \frac{v_{\Gamma}}{\Gamma} d\Gamma
\label{dngamma}
\end{eqnarray}
where $\frac{v_{\Gamma} d\Gamma }{\Gamma}$ represents the probability
decimated in a window of with $d\Gamma$ around $\Gamma$.
The factor $1/\Gamma$ represents the probability to be decimated
via a given exit channel near the infinite disorder fixed point,
 and thus the additional factor $v_{\Gamma}$
can be interpreted as an effective number of independent exit channels
that are in competition to be decimated.
For instance in the one-dimensional Sinai model, 
this factor is simply $v_{\Gamma}({1d})=2$ because there are exactly
2 independent neighbors at any stage of renormalization in $d=1$,
one on the left and one on the right. And this is why
the number of renormalized valleys decays as $1/\Gamma^2$.

On the other hand, if we use the asymptotic expression
of Eq. \ref{defpsi2} for the coherence length 
$l_{\Gamma} \opsimeq c \Gamma^{1/\psi}$, 
we obtain the following decay for the number
of surviving configurations of Eq. \ref{ngamma2}
\begin{eqnarray}
 d \ln { \cal N}_{\Gamma} = 
d \left( \frac{L}{ c \Gamma^{1/\psi}} \ln 2\right)
= - \frac{L}{ c \Gamma^{1/\psi}} \ \frac{ \ln 2}{\psi} \frac{d\Gamma}{\Gamma}
= - \frac{L}{l_{\Gamma} } \ \frac{ \ln 2}{\psi} \frac{d\Gamma}{\Gamma}
\label{dngamma2}
\end{eqnarray}
The identification of Eqs \ref{dngamma} and \ref{dngamma2}
yields that the effective number $v_{\Gamma}$
 of independent exit channels from a surviving configuration reads
\begin{eqnarray}
 v_{\Gamma}
=   \frac{ \ln 2}{\psi} \ \frac{L}{l_{\Gamma}}   =
\frac{ \ln 2}{\psi}  n_{\Gamma}
\label{vgamma}
\end{eqnarray}
It is proportional to the number  $n_{\Gamma} \equiv  \frac{L}{l_{\Gamma}}$
of renormalized degrees of freedom. 
Since the numerical prefactor $\frac{ \ln 2}{\psi}$ is finite for $\psi>0$,
this means that for a given renormalized degree,
the number of independent directions that are in competition
to be decimated remains effectively finite.

This discussion remains of course at a qualitative level,
since a complete characterization
of the random structure generated by the strong disorder
 RG flow remains a very challenging issue.
However it is important to understand the meaning of the
strong disorder RG procedure.
The full RG rules give at first sight the impression 
that the proliferation of neighbors could ruin the method.
We have seen that this is not the case, and that the RG flow
can still be towards an infinite disorder fixed point.
We have then explained how the preferred exit channel actually
dominates over the others asymptotically. And the present discussion
on the decay of the number ${\cal N}_{\Gamma}$ of 
renormalized configurations shows that for a given renormalized
degree of freedom of size given by the coherence length $l_{\Gamma}$,
the number of effective exit channels that are in competition
to be decimated is effectively finite and proportional to $1/\psi$.

\section{ Conclusion}

\label{conclusion}

In this paper, we have analyzed in details 
the strong disorder RG procedure in configuration space
to study the non-equilibrium dynamics of random systems.
In particular, we have shown that
whenever the flow of the renormalized barriers is
towards some ``infinite disorder'' fixed point,
the properties of the large time dynamics
can be obtained via simplified RG rules 
that are asymptotically exact,
because the preferred exit channel out of a given renormalized
 valley typically dominates over the other exit channels asymptotically.
 As an example of application,
we have followed numerically the RG flow for the case of a directed polymer
in a two-dimensional random medium.
The full RG rules have been used to check that the RG flow is towards
some infinite disorder fixed point, whereas the simplified RG rules
that allow to study bigger sizes have been used
to estimate the barrier exponent 
$\psi \sim 0.47$ of the fixed point, in reasonable agreement
with best numerical measure presently available $\psi \sim 0.49$
obtained via Langevin dynamics \cite{rosso}. 

From a theoretical point of view, 
we have explained why the RG flow towards an infinite disorder fixed point
in configuration space is a strong support to the 
droplet scaling theory  \cite{Fis_Hus}
where the dynamics is governed
by the logarithmic growth of the coherence length
 $l(t) \sim \left( \ln t \right)^{1/\psi}$,
and where two-times $(t_w,t_w+t)$ aging properties involve the ratio
$l(t_w+t)/l(t_w)$ of the coherence lengths, i.e. the ratio
$\ln (t_w+t)/ \ln (t_w)$.
Moreover, the statistics of barriers corresponds to a very strong hierarchy
of valleys within valleys, which is necessary to allow
the coexistence of rejuvenation and memory
effects in temperature cycling experiments \cite{bouchaud} : 
the rejuvenation due to short length scales
does not destroy the memory of large length scales 
which are effectively frozen.

Besides these aging properties, 
another important issue is the response of disordered
systems to an external force $F$.
This question is analyzed in detail in our recent work \cite{driven}
where we explain how the "infinite disorder fixed point" for $F=0$
becomes a "strong disorder fixed point" at small $F$
with an exponential distribution of renormalized barriers,
that leads to the existence of an anomalous zero-velocity phase
for the motion of driven interfaces in random media.

From a numerical point of view, it is clear that the formulation
of RG rules in configuration space has an exponential numerical price,
since the number of initial configurations ${\cal N}_0$
grows exponentially with the number of degrees of freedom, 
i.e. grows exponentially with the volume $L^d$ for a system of linear
size $L$ in dimension $d$.
This computational complexity is not surprising,
since the determination of barriers for the dynamics is 
expected to be an NP-complete problem \cite{middleton}.
For the case of a directed polymer in a two-dimensional random medium
considered in the present paper where ${\cal N}_0 =2^L$,
 we have been able to follow
the full RG rules up to $L \leq 11$, and the simplified RG rules
up to $L \leq 18$. So it is clear that the numerical study 
of higher dimensional disordered systems via strong disorder
RG rules requires other decisive improvements. The most promising idea is
to use the spatial locality of the dynamics and the fact that regions
separated by a distance bigger than the coherence length $l(t)$
are not yet dynamically correlated at time $t$.
This strategy of 'quasi-factorization' into patches of increasing
length scale has been successfully applied
recently in the context of Monte-Carlo exact sampling
of the two-dimensional Ising spin-glass \cite{chanal}. 
We thus hope that the numerical application of strong disorder RG 
in configuration space will become possible in the future
for two or higher dimensional disordered models.

\end{document}